\documentclass[twocolumn]{aastex631}
\usepackage{threeparttable}

\shorttitle{CO Emitters in SSA22-AzTEC26 Field}
\shortauthors{Huang et al.}
\graphicspath{{./}{figures/}}

\begin{document}
\title{Characterizing CO Emitters in the SSA22-AzTEC26 Field}

\author{Shuo Huang}
\affiliation{Institute of Astronomy, Graduate School of Science, The University of Tokyo, 2-21-1 Osawa, Mitaka, Tokyo 181-0015, Japan}
\affiliation{National Astronomical Observatory of Japan, 2-21-1 Osawa, Mitaka, Tokyo 181-8588, Japan}

\author{Hideki Umehata}
\affiliation{Institute for Advanced Research, Nagoya University, Furocho, Chikusa, Nagoya 464-8602, Japan}
\affiliation{Department of Physics, Graduate School of Science, Nagoya University, Furocho, Chikusa, Nagoya 464-8602, Japan}

\author{Ryohei Kawabe}
\affiliation{National Astronomical Observatory of Japan, 2-21-1 Osawa, Mitaka, Tokyo 181-8588, Japan}
\affiliation{Department of Astronomy, School of Science, The Graduate University for Advanced Studies (SOKENDAI), Osawa, Mitaka, Tokyo 181-8588, Japan}
\author{Kotaro Kohno}
\affiliation{Institute of Astronomy, Graduate School of Science, The University of Tokyo, 2-21-1 Osawa, Mitaka, Tokyo 181-0015, Japan}
\affiliation{Research Center for the Early Universe, Graduate School of Science, The University of Tokyo, 7-3-1 Hongo, Bunkyo-ku, Tokyo 113-0033, Japan}
\author{Minju Lee}
\affiliation{Cosmic Dawn Center (DAWN), Denmark}
\affiliation{DTU-Space, Technical University of Denmark, Elektrovej 327, DK2800 Kgs. Lyngby, Denmark}
\author{Yoichi Tamura}
\affiliation{Department of Physics, Graduate School of Science, Nagoya University, Furocho, Chikusa, Nagoya 464-8602, Japan}
\author{Bunyo Hatsukade}
\affiliation{Institute of Astronomy, Graduate School of Science, The University of Tokyo, 2-21-1 Osawa, Mitaka, Tokyo 181-0015, Japan}
\author{Ken Mawatari}
\affiliation{National Astronomical Observatory of Japan, 2-21-1 Osawa, Mitaka, Tokyo 181-8588, Japan}
\affiliation{Institute for Cosmic Ray Research, The University of Tokyo, 5-1-5 Kashiwanoha, Kashiwa, Chiba 277-8582, Japan}
\affiliation{Department of Environmental Science and Technology, Faculty of Design Technology, Osaka Sangyo University, 3-1-1, Nakagaito, Daito, Osaka, 574-8530, Japan}

\begin{abstract}
 We report the physical characterization of four CO emitters detected near the bright submillimeter galaxy (SMG) SSA22-AzTEC26. We analyze the data from Atacama Large Millimeter/submillileter Array band 3, 4, and 7 observations of the SSA22-AzTEC26 field. In addition to the targeted SMG, we detect four line emitters with a signal-to-noise ratio $>5.2$ in the cube smoothed with 300 km s$^{-1}$ FWHM Gaussian filter. All four sources have NIR counterparts within 1$\arcsec$. We perform UV-to-FIR spectral energy distribution modeling to derive the photometric redshifts and physical properties. Based on the photometric redshifts, we reveal that two of them are CO(2-1) at redshifts of 1.113 and 1.146 and one is CO(3-2) at $z=2.124$. The three sources are massive galaxies with a stellar mass $\gtrsim10^{10.5}M_\odot$, but have different levels of star formation. Two lie within the scatter of the main sequence (MS) of star-forming galaxies at $z\sim1-2$, and the most massive galaxy lies significantly below the MS. However, all three sources have a gas fraction within the scatter of the MS scaling relation. This shows that a blind CO line search can detect massive galaxies with low specific star formation rates that still host large gas reservoirs and that it also complements targeted surveys, suggesting later gas acquisition and the need for other mechanisms in addition to gas consumption to suppress star formation.
\end{abstract}

\section{Introduction}\label{sec:intro}
The cosmic star formation rate (SFR) density increases from early times to its peak at $z\sim2$, called cosmic noon, then decreases progressively to the present day \citep[for a review, see][]{2014ARA&A..52..415M}. Across cosmic history, the most massive galaxies  (stellar mass $M_\star\gtrsim10^{11}M_\odot$) have formed the bulk of their $M_\star$ around or earlier than the cosmic noon and ceased star formation in late epochs \citep[e.g.,][]{2010MNRAS.404.1775T,2013ApJ...777...18M,2014ApJ...783...85T,2015MNRAS.448.3484M,2017A&A...605A..70D}, while the formation of less massive galaxies continues for a longer period. Molecular gas is a key factor in shaping the history of galaxy assembly, as it is the immediate material of star formation \citep[for a review, see][]{2013ARA&A..51..105C}. It has been suggested that galaxies acquire molecular gas via accretion from the intergalactic medium \citep[e.g.,][]{2009Natur.457..451D,2015Natur.525..496N} or mergers to fuel their star formation and central black hole growth \citep[e.g.,][]{2008ApJS..175..356H}. Observations of molecular gas content in local and distant galaxies thus provide important clues about galaxy formation and evolution.

\par
CO rotational transition lines are commonly used to trace the molecular gas content of galaxies because molecular hydrogen is a poor emitter and CO is the second most abundant molecule in the interstellar medium (ISM). Numerous observations of CO line emission have been conducted to obtain statistical samples of galaxies at intermediate and high redshifts ($z\sim1-3$) to study the relation between gas and other properties of galaxies, such as $M_\star$ and SFR. CO observations of high-redshift galaxies typically select massive normal star-forming galaxies \citep[SFGs; e.g.,][]{2010ApJ...714L.118D,2010Natur.463..781T} or the most extreme starbursting galaxies \citep[e.g.,][]{2003ApJ...597L.113N,2010ApJ...724..233E,2013MNRAS.429.3047B}. These surveys have successfully established scaling relations that describe how galaxy properties evolve with the gas mass \citep[e.g.,][]{2015ApJ...800...20G,2018ApJ...853..179T,2020ARA&A..58..157T}. 

\par
Observations of a large sample of galaxies have revealed a tight correlation between $M_\star$ and SFR called the main sequence \citep[MS; e.g., ][]{2007A&A...468...33E,2010ApJ...714L.118D,2011ApJ...742...96W,2012ApJ...754L..29W,2014ApJS..214...15S,2015A&A...575A..74S}. SFGs on the MS dominate the cosmic star formation \citep{2011ApJ...739L..40R,2012ApJ...747L..31S}, implying the existence of mechanisms that regulate star formation from gas and the steady buildup of galaxies \citep[e.g.,][]{2013ApJ...772..119L}. Below the scatter of the MS, there is another population of passive or quiescent galaxies (QGs), with little or no ongoing star formation \citep[e.g., ][]{2001AJ....122.1861S,2003MNRAS.341...54K,2004ApJ...600..681B,2015ApJS..219....8C}. Many physical processes have been proposed to explain the shutdown of star formation, including the starvation of gas, either due to the removal of cold gas by stellar and supermassive black hole feedback \citep[e.g.,][]{2006ApJS..163....1H,2010MNRAS.401....7H} or rapid gas consumption by vigorous star formation  \citep[e.g.,][]{1999ApJ...512L..99G,2019A&A...624A..81M}, and the inability of the conversion of gas into stars due to changes in the ISM condition \citep{2018ApJ...861..123F} or stabilization against collapse and fragmentation \citep[e.g.,][]{2009ApJ...707..250M}.
In order to understand the roles of these processes in galaxy quenching, it is essential to characterize the gas properties such as the gas fraction, depletion timescale, and kinematics.

\par
In the case of massive galaxies, this requires detecting CO emission at $z\gtrsim1$, which corresponds to the epoch when many of them are undergoing formation and subsequent quenching. Most of the aforementioned high-redshift CO observations target preselected MS galaxies or starbursts, thus current scaling relations \citep[e.g.,][]{2018ApJ...853..179T} only account for SFGs. Whether these relations are still valid when extrapolated to passive galaxies remains unknown. Recent targeted observations of CO emission from QGs have shown a mixed picture \citep[e.g.,][]{2021ApJ...909L..11B,2021ApJ...908...54W}. An unbiased CO survey of galaxies in various stages of star formation is needed to investigate the evolution of gas content before and after galaxy quenching. By detecting both low and medium to high $J$ transitions, it is also possible to study the redshift evolution of the molecular gas in galaxies. Several studies have used the Atacama Large Millimeter/submillimeter Array and JVLA to perform blind line searches toward deep cosmological fields or ALMA calibrators \citep{2016ApJ...833...67W,2019ApJ...882..139G,2019ApJ...872....7R,2020ApJ...896L..21R,2023MNRAS.519...34H}. However, because of the large amounts of telescope time needed, the surveyed area and the number of detected sources are still very limited. 

\par
In this paper, we present serendipitous detections and physical properties of four CO emitters in the vicinity of the bright submillimeter galaxy (SMG)  SSA22-AzTEC26 from ALMA band 3 observations. In Section \ref{sec:sample} we describe the ALMA observation and data analysis. In Section \ref{sec:result} we present the extracted galaxy properties utilizing multiwavelength data in the SSA22 field. In Section \ref{sec:discussion}, we discuss the gas content of CO-selected galaxies in the context of galaxy evolution, and then we summarize in Section \ref{sec:summary}.

\par
Throughout this paper, we adopt the \citet{2003PASP..115..763C} initial mass function and Planck 2018 cosmology \citep[$H_0=67.4$ km s$^{-1}$ Mpc$^{-1}$, $\Omega_{m}=0.315$ and $\Omega_{\mathrm{\Lambda}}=0.685$; ][]{2020A&A...641A...1P}. 

 \section{Sample and data analysis}\label{sec:sample}
 
\subsection{CO Emitters from ALMA Band 3 Observations}\label{subsec:sample}

\subsubsection{ALMA data}\label{subsubsec:almaobs}
SSA22-AzTEC26 is a bright SMG first discovered by the AzTEC/ASTE 1.1 mm extragalactic survey in the SSA22 field \citep{2009Natur.459...61T,2014MNRAS.440.3462U}. The ALMA band 3 spectral scan observations (Project ID: 2019.1.01102.S; PI: Umehata) were conducted from 2020 March 20 to 2020 April 3, toward the sky position of SSA22-AzTEC26 (R.A. 22:17:13.34, decl. 00:26:51.66). The observations had a maximum baseline of 313.7 m and continuously covered the sky frequency range from 84.5 to 113.7 GHz with five tunings. The total integration time was 6.2 hr. J2217+0220 and J2206-0031 were observed for phase calibration. To calibrate the flux and bandpass, J0006-0623, J2253+1608, J2258-2758, and J1924-2914 were observed.

\par
We use the \texttt{CASA} package \citep[][]{2022PASP..134k4501C} to reduce the data and perform imaging. A spectral cube is created using the \texttt{tclean} task with natural weighting and a 100 km s$^{-1}$ channel width. The resulting band 3 spectral cube has a $\sim1\farcm5$  arcmin diameter field of view (FOV) with a minimum primary beam (PB) response of 20\% of the field center. The synthesized beam size is $3\farcs80\times2\farcs45$ with a position angle (PA) of $57\fdg7$ to $4\farcs38\times3\farcs65$ awith PA of $81\fdg2$ degrees, depending on the sky frequencies. The RMS noise level range is $0.06-0.18$ mJy beam$^{-1}$ before PB correction.

\par
We also include band 4 and band 7 data (Project ID: 2021.1.01207.S; PI: Umehata) in our analysis. The band 4 observations cover the frequency ranges of 143.0-146.7 GHz and 154.9-158.7 GHz with a single tuning and total integration time of 2.5 hr. We process the data in a similar manner as for band 3. The resulting band 4 cube has a synthesized beam size of $0\farcs87\times0\farcs79$ to $0\farcs96\times0\farcs88$ with PA of $\sim61$ degrees, and RMS noise level of $0.05-0.09$ mJy beam$^{-1}$ in a $\sim40\arcsec$ FOV. The band 7 data cover the frequency range 340.4-356.2 GHz with the beam size of $0\farcs63\times0\farcs52$ with a PA of $88\fdg8$ and an RMS noise of 0.09 mJy beam$^{-1}$ in the 860 $\mu$m continuum map.

\subsubsection{Source detection and line measurements}\label{subsubsec:detection}
\par
\begin{figure*}[!htp]
    \plotone{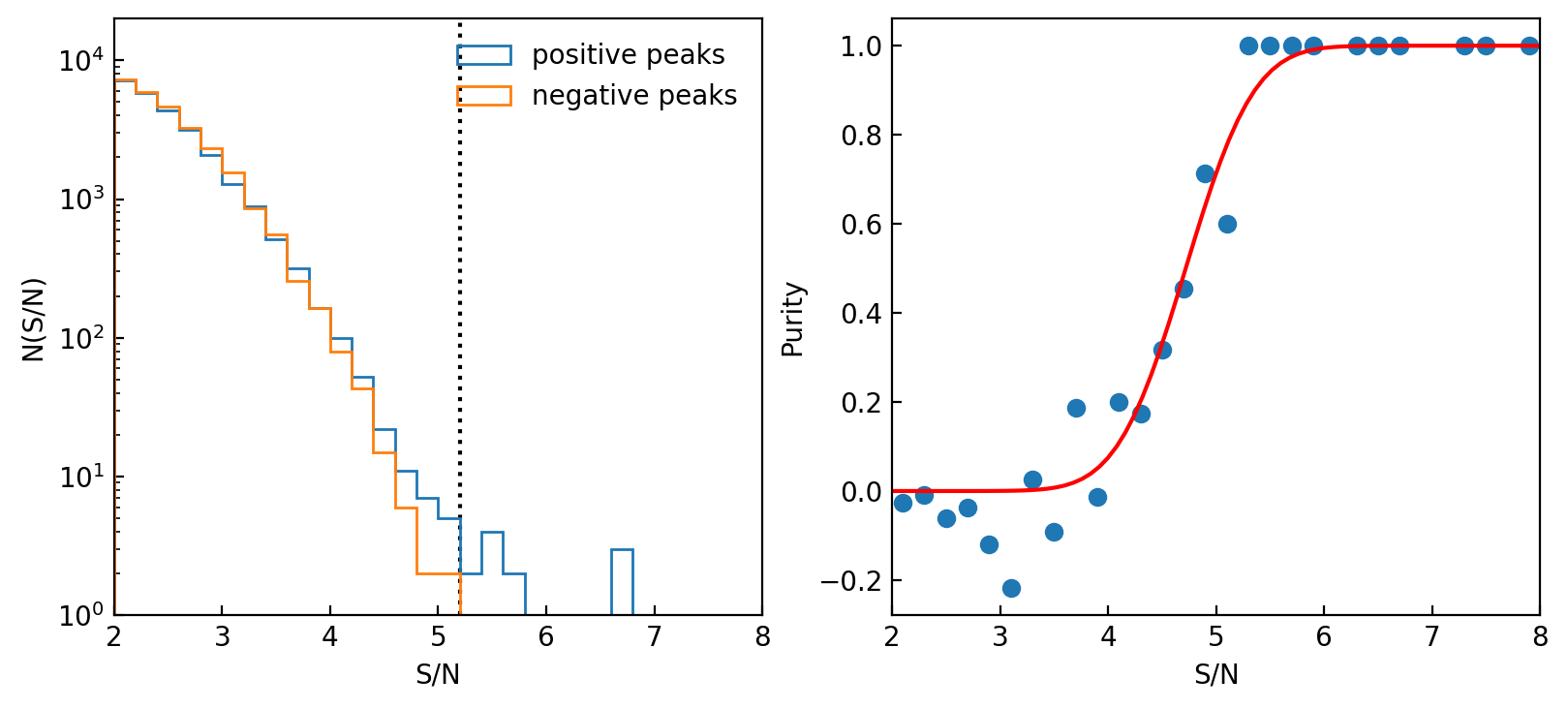}
    \caption{Left: the number of detections as a function of positive or negative peak S/N. The S/N$>5.2$ detection threshold is marked as a vertical dotted line. Right: purity as a function of peak S/N. The best-fit relation of Equation \ref{eq2} is shown as a red curve.}
    \label{fig:fig1}
\end{figure*}
\begin{figure}[ht]
    \plotone{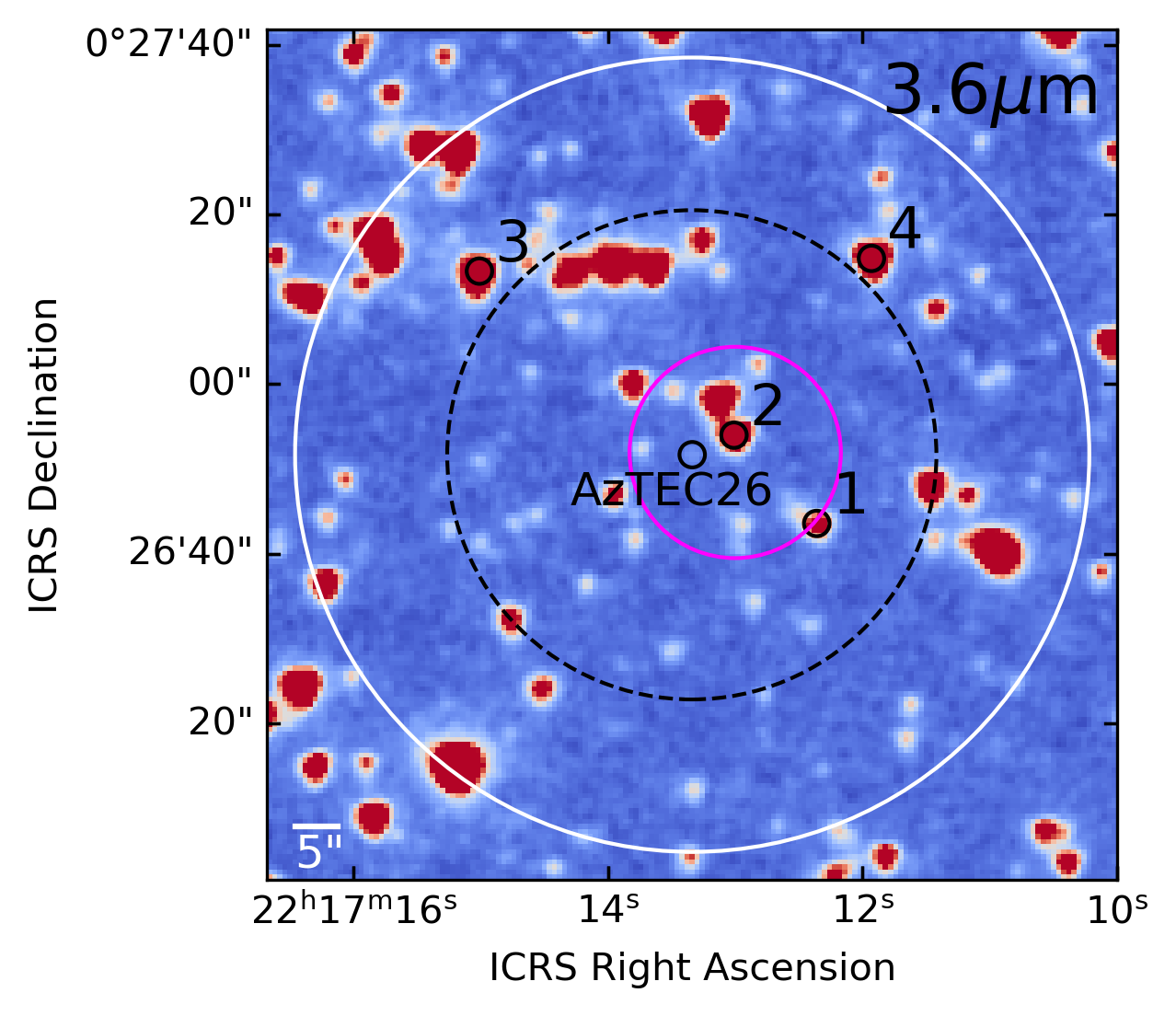}
    \caption{IRAC 3.6 $\mu$m image of the targeted field. The sky locations of SSA22-AzTEC26 and the four line emitters are marked with black empty circles. { The ALMA FOVs with $20\%$ PB response in bands 3, 4, and 7 are shown as the white, black dashed, and magenta circles, respectively.}}
    \label{fig:fig2}
\end{figure}

Inspired by \citet{2016ApJ...833...67W}, we smooth the band 3 cube with a 300 km s$^{-1}$ FWHM Gaussian filter and use the \texttt{DAOStarFinder} routine in the \texttt{photutils} package to detect all positive and negative peaks above a peak signal-to-noise ratio (S/N) of 1.5 in the smoothed cube. The central AzTEC source is masked because its continuum emission is visible in the smoothed cube and contaminates source statistics. We show the number of detected peaks as a function of peak S/N in the left panel of Figure \ref{fig:fig1}. No negative peak is found above S/N$=5.2$. 
The purity of detection as a function of S/N is defined as
\begin{equation}\label{eq1}
    \mathrm{Purity} = 1-\frac{N_\mathrm{negative}}{N_\mathrm{positive}}.
\end{equation}
We fit the measured purity as a function of S/N with
\begin{equation}\label{eq2}
    P(\mathrm{S/N})=\frac{1}{2}\mathrm{erf}\left(\frac{\mathrm{S/N}-a}{b}\right)+0.5,
\end{equation} and determine the free parameters $a=4.72$ and $b=0.71$, which gives a purity of 83\% at S/N=5.2. Using this as a detection threshold, we find four sources with S/N$>5.2$ (Figure \ref{fig:fig2}). We have also tested other filter FWHM values from 100 to 500 km s$^{-1}$ with 100 km s$^{-1}$ steps but do not find any new detection.

\par
For each source, we fit a single Gaussian profile to the 1D spectrum to derive the observed line frequency and FWHM (the bottom row in Figure \ref{fig:fig3}). The velocity-integrated fluxes are measured by fitting a 2D Gaussian profile to the zeroth-moment map, which sums along the spectral axis over a velocity range $\pm$FWHM from the line center. The measured properties are listed in Table \ref{table:table1}. Sources 1, 2, and 4 are covered by the band 4 observation, so we also report the 2 mm continuum flux densities for these three sources.  Source 1 has a second line detection at 157.548 GHz in the band 4 cube, and we will show that this is the redshifted [CI] line in the next section. 

\begin{deluxetable*}{cccccccccc}
\tablenum{1}
\caption{\label{table:table1}Results of Blind Line Emitter Search in the Band 3 Cube and Source Measurements.}
\tablewidth{0pt}
    \tablehead{\colhead{ID} & \colhead{R.A.} & \colhead{Decl.} &\colhead{$\nu_{\mathrm{obs}}$} & \colhead{S/N} & \colhead{FWHM} & \colhead{$S\Delta v$} &\colhead{$S_\mathrm{3mm}$} &\colhead{$S_\mathrm{2mm}$} & \colhead{$S_{860\mu\mathrm{m}}$}\\
    \colhead{} & \colhead{} & \colhead{} & \colhead{(GHz)} & \colhead{} & \colhead{(km s$^{-1}$)} & \colhead{(Jy km s$^{-1}$)} & \colhead{($\mu$Jy)} & \colhead{($\mu$Jy)} & \colhead{(mJy)}}
         \startdata
         1&22:17:12.36 &00:26:43.59&110.698  & 5.9&697$\pm$153 &0.569$\pm$0.075&$14.8\pm2.4$ &$90\pm18$ & -\\ 
         2&22:17:13.00 &00:26:54.06&109.151 &7.8 &635$\pm$137&0.358$\pm$0.036&$<13$ &$39\pm7.3$ & $0.815\pm0.194$\\
         3&22:17:15.01 &00:27:13.32&91.785 &7.3 &252$\pm$85&0.450$\pm$0.075& $21.7\pm3.1$ & - & -\\
         4&22:17:11.93 &00:27:14.82&107.414 &5.5 &630$\pm$123&0.713$\pm$0.141&$<27$ & $<100$ & - \\
        \enddata
\tablecomments{The 3mm fluxes and the 2 mm flux of source 1 are measured using line-free spectral windows. For non-detection, we report $3\sigma$ upper limit. Source 1 has another line detection at 157.548 GHz, with $S\Delta v=0.320\pm0.044$ Jy km s$^{-1}$ and FWHM of $683\pm201$ km s$^{-1}$.}
\end{deluxetable*}

\subsection{Ancillary Data}\label{subsec:ancdata}
\subsubsection{Optical and NIR data}\label{subsubsec:optnir}
\begin{figure*}[ht]
    \centering
    \includegraphics{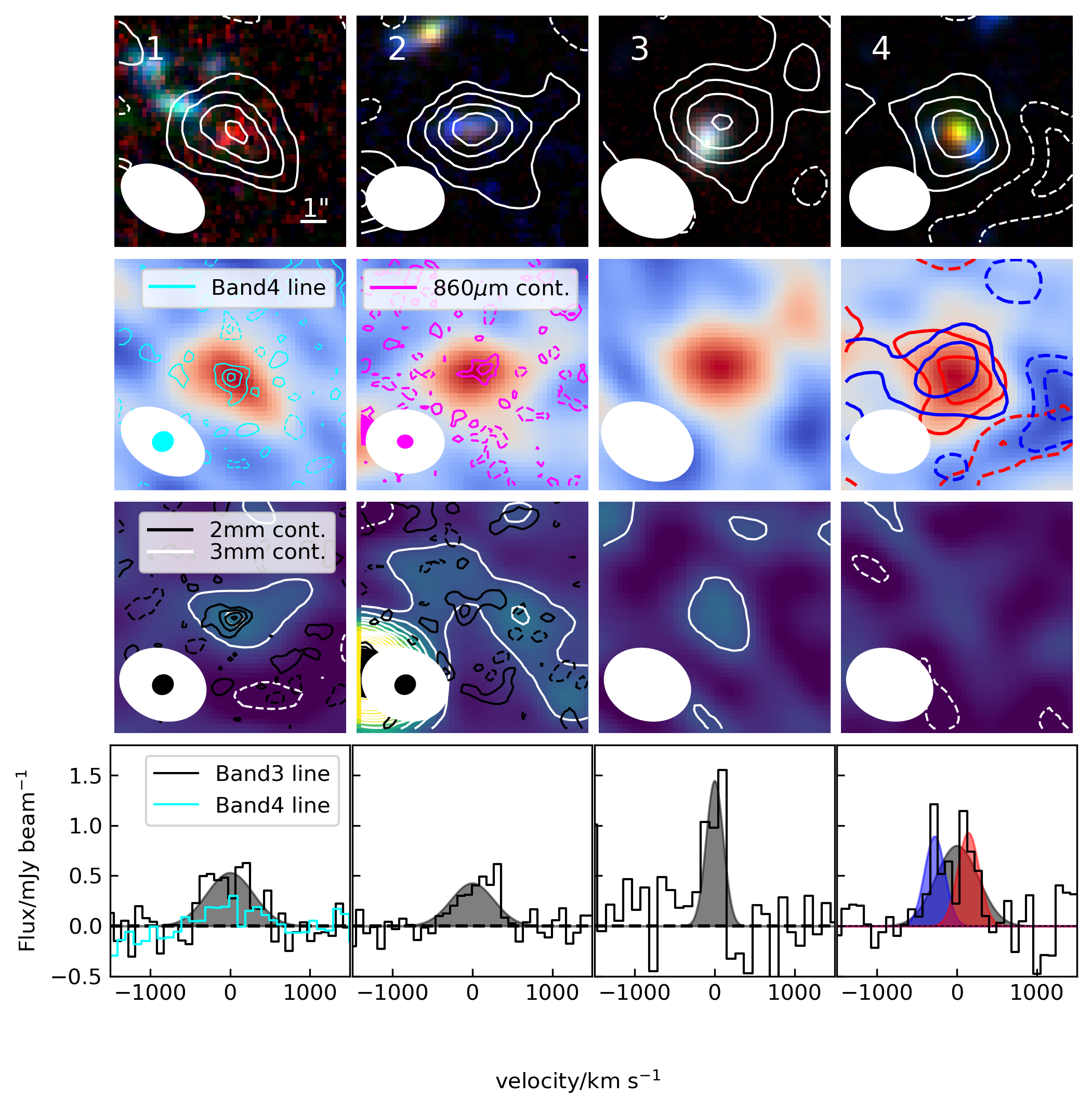}
    \caption{
    Top row: color-composite (red: $K$; green: $z'$; blue: $B$) cutout images of the line emitters. Each image is $10\arcsec\times10\arcsec$ in size. The zeroth-moment maps of the band 3 emission line are overlaid as white contours. The corresponding beam sizes are shown with white ellipses.
    Second row: the zeroth-moment maps of the band 3 emission line.  The zeroth-moment map of the band 4 157.548 GHz emission line of source 1 is shown with cyan contours. For source 2 we show the band 7 continuum image with magenta contours. All contours start from $\pm1.5\sigma$ with a step of $1.5\sigma$, and the corresponding beam sizes are shown with the same colors.
    Third row: band 3 continuum images are shown with background and white contours. The band 4 images of sources 1 and 2 are overlaid as black contours.
    Bottom row: 1D band 3 spectra at the position of each source. The best-fit single Gaussian model is shown with a black shaded region. For source 1, the band 4 emission line is shown in cyan. For source 4, we fit two Gaussian components and show the corresponding zeroth-moment contour in the second row with the same color.}
    \label{fig:fig3}
\end{figure*}

The SSA22 field has rich multiwavelength data coverage. We collect ground-based optical-to-NIR images from the $u$ to $K$ bands from archives and literature. The thumbnails of the sources are shown in the top row of Figure \ref{fig:fig3}. The astrometry is corrected by stacking cutout images at the positions of 1.1 mm continuum sources detected by ALMA in the SSA22 field (B. Hatsukade, private communication) and fitting a 2D circular Gaussian profile to the stacked image to determine the average positional offset between the ALMA source and its counterpart in each band. Sources 1, 2, and 4 have a coincident $K$-band counterpart with a centroid offsets of $0\farcs27$, $0\farcs23$, and $0\farcs14$, respectively, comparable with a position uncertainty of $\sim0.2\arcsec$ of the line detection. The NIR counterpart of source 3 lies $0\farcs62$ southeast of the CO position, $\sim$three times of the uncertainty of the line position.

\par
\begin{figure}
    \plotone{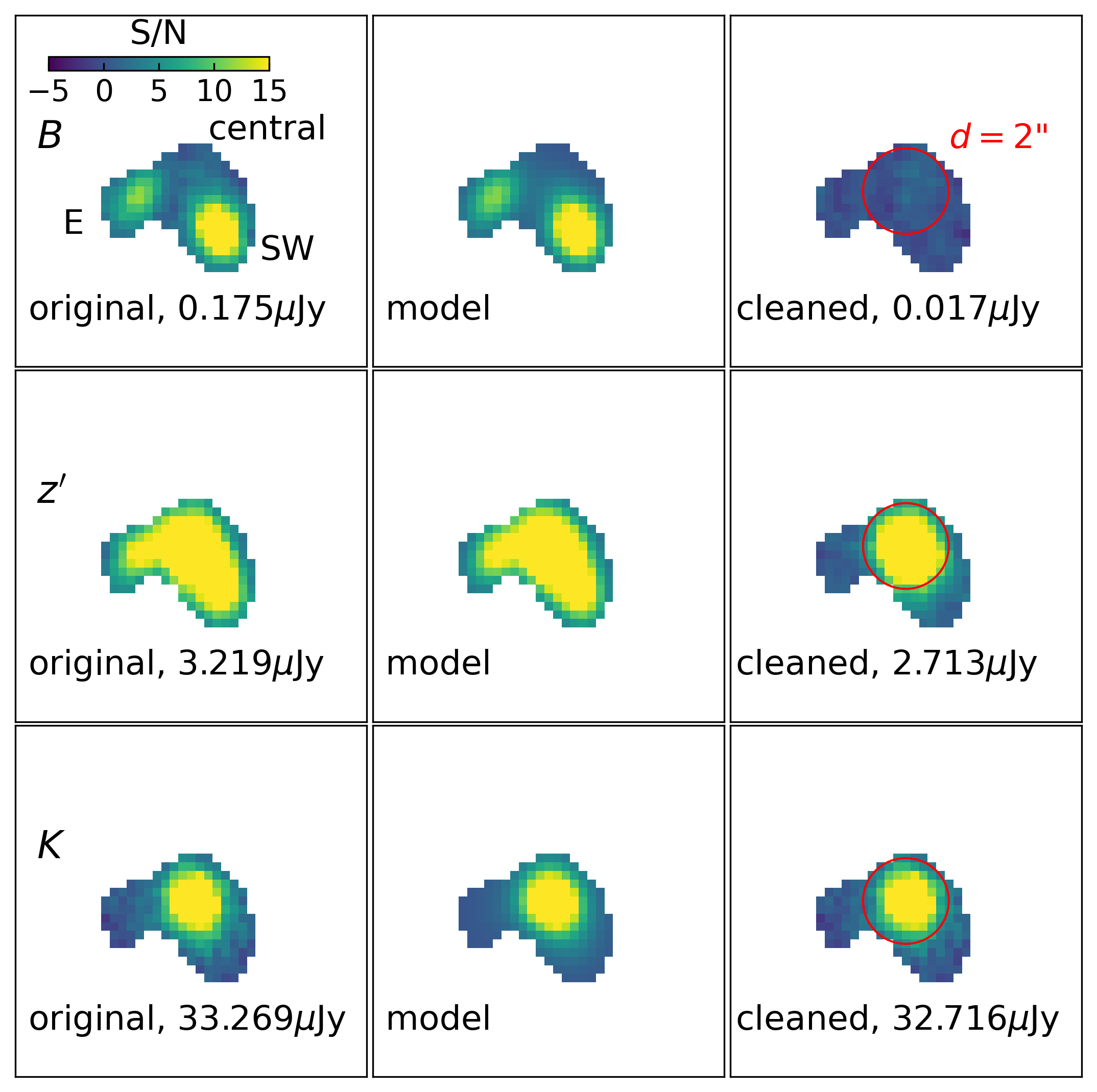}
    \caption{Examples of the neighbor subtraction of galaxy 4. The left, middle, and right columns show original data, the best-fit Sersic model, and the neighbor-subtracted data, respectively. The $2\arcsec$ diameter aperture used to measure the flux is shown as a red circle. The relationship between the three components will be further discussed in section \ref{subsubsec:gal4clump}.}
    \label{fig:fig4}
\end{figure}

We measure the flux densities using a $2\arcsec$ diameter circular aperture placed at the CO position in bands shorter than IRAC. The errors are estimated by taking the standard deviations of fluxes from 1000 randomly placed apertures at blank positions. Then we aperture correct the fluxes using the point spread function (PSF) and correct for galactic extinction using the \citet{1998ApJ...500..525S} dust map and \citet{1989ApJ...345..245C} Milky way extinction curve. 
{
PSFs are generated using the \texttt{PSFEx} \citep{2011ASPC..442..435B} code from unsaturated point sources with S/N$\geq20$.
\par
As shown in Figure \ref{fig:fig3}, the aperture flux of source 4 is contaminated by two blue neighboring sources.  At this moment, we simply assume the two blue sources are foreground galaxies and subtract their contributions using 2D image modeling. We use a single S\'{e}rsic profile convolved with the PSF to model each galaxy and then perform Bayesian inference using the \texttt{dynesty} code \citep{2020MNRAS.493.3132S} to derive posterior probability density functions (PDFs) of the model parameters. Flat priors of [0, 6] and [0.1, $5a_\mathrm{image}$] pixels are adopted for the S\'{e}rsic index $n$ and half-light radius $R_\mathrm{e}$, respectively, where $a_\mathrm{image}\sim2$ pixels is the semi-major axis sigma value given by \texttt{photutils} and the pixel size is $0\farcs2$. We first model the central source using the $K$-band image and then the two neighboring galaxies jointly using the $B$-band image to minimize the effect of blending. In subsequent subtraction processes, the Sersic shape parameters are limited to the 16th to 84th percentile range from the previous procedure and the central position can vary by $\pm0.5$ pixels from the median. The source flux is always allowed to vary freely. Then we simultaneously fit all three sources in each band from $u$ to $K$. When measuring the flux of neighboring sources, we subtract the central source, and for the central source, we subtract the two neighboring clumps. Deblended fluxes and errors are derived from the posterior PDF in each band, with formal photometric errors added in quadrature. Examples of subtracting processes are demonstrated in Figure. \ref{fig:fig4}. Further discussion of the relation between the central and neighboring blue sources is given in section \ref{subsubsec:gal4clump}.
}

\subsubsection{Infrared data}\label{subsubsec:irdeblending}
 For long-wavelength NIR (3.6 $\mu$m) to MIR, we obtain IRAC 3.6-8 $\mu$m and MIPS 24$\mu$m images from the Spitzer archive. At FIR wavelengths, we use Herschel SPIRE 250-500 $\mu$m images from the ESA Herschel archive. Flux densities are derived from PSF photometry to overcome source blending issues. For the IRAC data, we apply the method introduced by \citet{2012ApJS..203...23H} which iteratively decomposites the image into a combination of ideal point sources and background noise. 
 We consider decomposed point sources within $1\arcsec$ (1.5 pixels of IRAC images) from the CO position as associated with the target and use the sum of their amplitudes as the total flux. Errors are estimated from the RMS of the residual image.
 \par
 In the Spitzer/MIPS and Herschel/SPIRE bands, we measure the flux densities by directly fitting PSF models to the image with IRAC 4.5 $\mu$m and 1.1 mm source positions as priors, following the method of \citet{2017MNRAS.464..885H}. Since our sources lie on the edge of the MIPS coverage, we also add sources detected at S/N$>5$ the in IRAC 8 $\mu$m images when deblending the SPIRE images to add dusty sources that are not covered by MIPS but might appear in SPIRE images. Starting from MIPS, we create a $2\arcmin\times2\arcmin$ cutout image centered at the line position. For each position $(x_i,y_i)$ in all prior sources, the source model $S_{i}(x,y)$ is made by interpolating the PSF centered at $(x_i,y_i)$ to the pixel grid $(x,y)$ and scaling it to unity flux density. The $\chi^{2}$ can be expressed as
 \begin{equation}
     \chi^{2} = \sum_{x,y}\left(\frac{I(x,y)-\sum_{i}f_{i}S_{i}(x,y)}{\sigma(x,y)}\right)^{2},
 \end{equation}
 where $I(x,y)$ is the data and $\sigma(x,y)$ is the flux uncertainty at each pixel. Then we perform Bayesian inference to derive the posterior PDF of the source flux densities $f_i$. Only sources that have a deblended S/N$>1$ will be used to fit the next band. None of the sources is detected at S/N$>3$ in any of the MIR/FIR images. The measured fluxes are listed in Table \ref{table:table2}.

\begin{deluxetable*}{ccccccc}[ht]
 \tablenum{2}
\caption{\label{table:table2}Multiwavelength Photometry of the Four CO Emitters in Units of $\mu$Jy, Extracted at the CO Position.}
\tablewidth{0pt}
\tablehead{\colhead{Band} & \colhead{Instrument} &\colhead{1} &\colhead{2} &\colhead{3} &\colhead{4} &\colhead{Reference}}
\startdata
$u$&CFHT/MegaCam &...&$0.125\pm0.020$&$1.183\pm0.019$&$0.078\pm0.020$&1 \\
$B$&Subaru/SuprimeCam &$0.057\pm0.006$&$0.113\pm0.007$&$1.388\pm0.007$&$0.018\pm0.009$&2 \\
$V$&- &$0.109\pm0.009$&$0.199\pm0.010$&$2.303\pm0.009$&$0.106\pm0.010$&2 \\
$R$&- &$0.142\pm0.014$&$0.371\pm0.013$&$3.316\pm0.013$&$0.346\pm0.018$&3 \\
$i'$&- &$0.186\pm0.018$&$0.748\pm0.018$&$5.853\pm0.020$&$0.911\pm0.024$&2 \\
$z'$&- &$0.311\pm0.034$&$1.609\pm0.036$&$8.122\pm0.038$&$2.772\pm0.042$&2 \\
NB359&- &$0.067\pm0.019$&$0.058\pm0.020$&$1.223\pm0.019$&$0.131\pm0.020$&4 \\
NB497&- &$0.099\pm0.014$&$0.124\pm0.015$&$1.874\pm0.015$&$0.082\pm0.017$&5 \\
NB816&- &$0.249\pm0.036$&$1.023\pm0.037$&$6.455\pm0.036$&$1.336\pm0.044$&2 \\
NB912&- &$0.265\pm0.029$&$1.462\pm0.030$&$7.906\pm0.029$&$3.064\pm0.038$&2 \\
$g$&Subaru/HSC &$0.083\pm0.017$&$0.133\pm0.016$&$1.828\pm0.017$&$0.089\pm0.020$&6 \\
$r$&- &$0.115\pm0.031$&$0.280\pm0.033$&$3.077\pm0.034$&$0.321\pm0.043$&6 \\
$i$&- &$0.177\pm0.027$&$0.793\pm0.030$&$6.245\pm0.030$&$0.990\pm0.038$&6 \\
$z$&- &$0.284\pm0.079$&$1.668\pm0.076$&$7.797\pm0.077$&$2.828\pm0.098$&6 \\
$Y$&- &...&$2.123\pm0.126$&$8.042\pm0.123$&$4.289\pm0.146$&6 \\
$J$&UKIRT/WFCAM &$1.019\pm0.206$&$4.137\pm0.214$&$13.052\pm0.214$&$8.458\pm0.241$&7 \\
$K$&- &$4.071\pm0.322$&$13.838\pm0.310$&$25.458\pm0.340$&$32.887\pm0.388$&7 \\
IRAC1&Spitzer/IRAC &$11.565\pm0.220$&$25.925\pm0.192$&$44.430\pm0.704$&$67.229\pm0.556$&8,9 \\
IRAC2&- &$18.926\pm0.851$&$25.833\pm1.121$&$44.152\pm0.815$&$59.478\pm2.734$&8,9\\
IRAC3&- &$17.952\pm2.103$&$17.459\pm1.747$&$40.255\pm1.980$&$37.444\pm2.027$&8,9 \\
IRAC4&- &$15.686\pm2.269$&$18.327\pm2.857$&$28.632\pm2.483$&$32.981\pm3.147$&8,9\\
MIPS1&Spitzer/MIPS &$305\pm149$&$20\pm44$&$392\pm169$&$115\pm189$&8,9 \\
PSW&Herschel/SPIRE &$6133\pm4584$&$4931\pm4861$&$4997\pm5191$&$3432\pm4442$&10,11 \\
PMW&- &$6437\pm7227$&$11129\pm6856$&$5741\pm6574$&$2406\pm4033$&10,11 \\
PLW&- &$1941\pm3710$&$4385\pm5905$&$2523\pm4818$&$1284\pm2745$&10,11 \\
\enddata
\tablerefs{
1: K. Mawatari, private communication;
2: \citet{2011MNRAS.412.2579N};
3: \citet{2004AJ....128.2073H};
4: \citet{2009ApJ...692.1287I};
5: \citet{2012AJ....143...79Y};
6: \citet{2018PASJ...70S...8A} 
7: \citet{2007MNRAS.379.1599L}; 
8: \citet{2009ApJ...692.1561W};
9: \citet{SpitzerHeritageArchive}
10: \citet{2016MNRAS.460.3861K};
11: \url{http://herschel.esac.esa.int/Science_Archive.shtml}
}
\tablecomments{``-" means the same as the previous band.}
\end{deluxetable*}

\subsection{Spectral Energy Distribution Modeling}\label{subsec:SEDfitting}
We adopt the methodology of \texttt{prospector} \citep{2019ApJ...877..140L} to model the observed spectral energy distribution (SED), which fits all physical parameters simultaneously and yields the joint PDF as a result. We model the dust attenuation for old and young stars with two \citet{2009A&A...507.1793N} dust attenuation curves, which are the \citet{2000ApJ...533..682C} dust extinction curve multiplied by a power-law modification $(\lambda_\mathrm{rest}/550\mathrm{nm})^{\beta}$. For the two stellar populations separated by age, the slope $\beta$ and amplitude of each curve can vary independently within a flat prior [-0.7, 0.4] for $\beta$ and [0, 10] for $E(B-V)$, while the dust attenuation of old stars is forced to be smaller than that of young stars. This adds some additional flexibility in  taking possible complex dust geometry into account. To implement the modification, SED models are built with the \texttt{CIGALE} code \citep{2019A&A...622A.103B}. Before line identification, we let the redshift vary freely within a flat prior [0, 8] (i.e., the photometric redshift mode of SED fitting).

\section{Results}\label{sec:result}
\subsection{Line Identification}\label{subsec:zsolve}
\begin{figure*}[htp!]
    \plotone{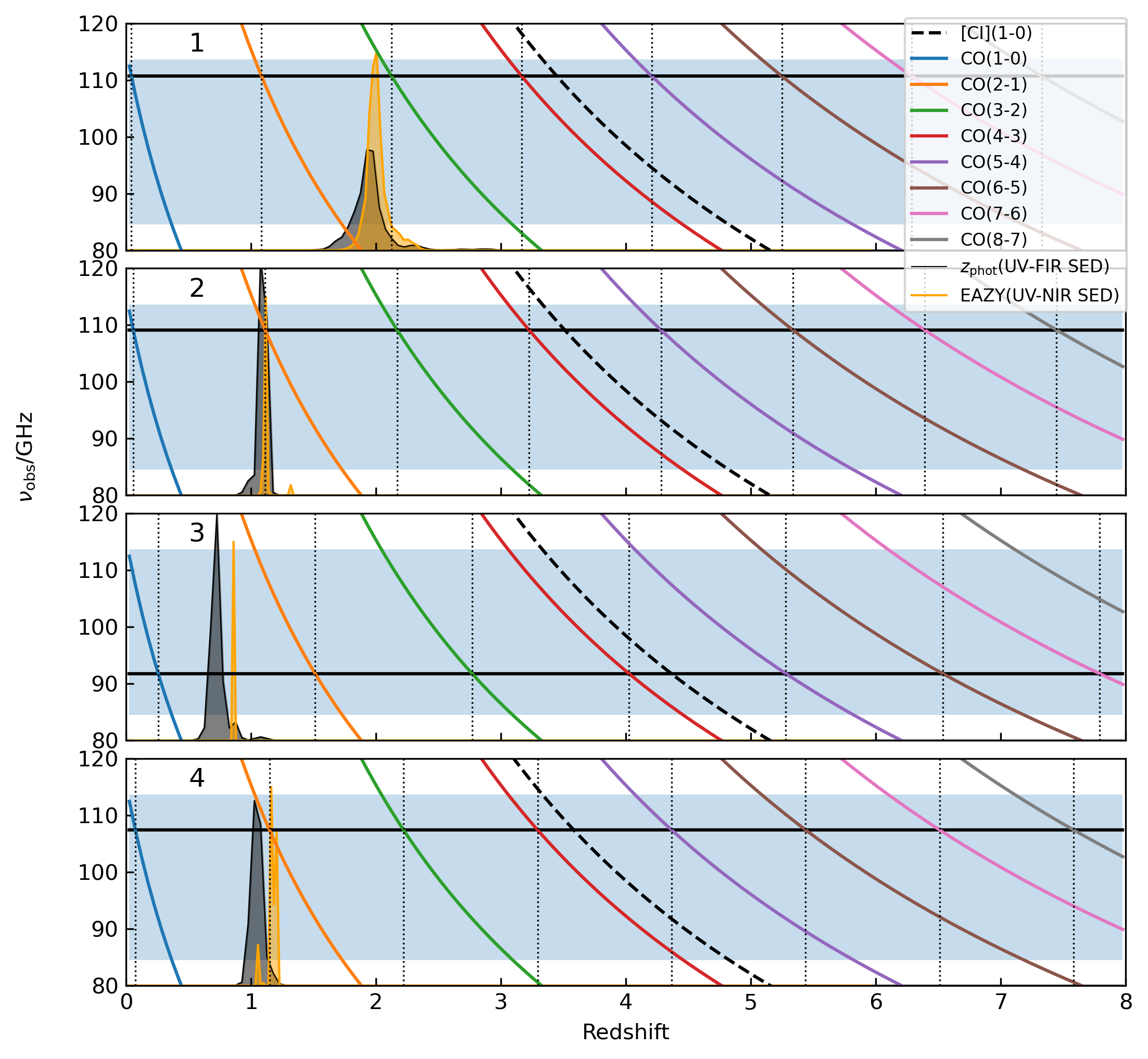}
    \caption{Observed frequency as a function of redshift for different spectral lines. CO lines are shown with colored curves and the [CI] line is shown with the dashed black curves. The blue shaded regions show the frequency coverage of the ALMA band 3 observations. We mark the observed frequency of each galaxy with a horizontal black line and the corresponding redshift of each solution with a vertical dotted line. The PDFs of photometric redshift are illustrated by lines and shaded regions, with black color for the SED fitting method described in Section \ref{subsec:SEDfitting} and orange color for \texttt{EASY}. A successful solution requires the vertical line to locate within the photometric redshift PDF.}
    \label{fig:fig5}
\end{figure*}

Each of the sources has only one robust line detection in the band 3 cube. The most probable identification of the detected line will be one of CO $J_\mathrm{up}=4,3,2,1$ from galaxies at $z\lesssim3-4$ because our $\sim29$ GHz bandwidth will cover two CO lines or one CO and [CI]($^3\mathrm{P}_1-^3\mathrm{P}_0$) (hereafter, [CI](1-0)) atomic carbon emission line for $z\gtrsim3-4$. To find the redshift solution, we plot the marginalized PDF of the photometric redshift from our SED fitting and \texttt{EAZY}\footnote{\url{https://github.com/gbrammer/eazy-py}} code \citep{2008ApJ...686.1503B}, together with the observed frequency as a function of redshift for the CO lines and [CI](1-0) line in Figure \ref{fig:fig5}.

\par
Based on the results of SED fitting, we find that all four sources are distant galaxies at a cosmological distance.  For galaxies 1, 2, and 4, there is a unique solution for CO line redshift that is allowed by the photometric redshift PDF. The redshift of 2.124 given by the CO J=3-2 line of galaxy 1 agrees with the second line in band 4 being redshifted [CI](1-0), which yields a consistent redshift $z_{\mathrm{[CI](1-0)}}=2.124$.  The photometric redshift PDF of galaxy 3 is incompatible with all solutions. Since the flux is extracted at the CO position, we check the results by changing the position of the photometry aperture to the center of the nearby NIR object, then the photometric redshift of $0.63^{+0.07}_{-0.06}$ still does not overlap with any solution. Thus, the flux is likely to be dominated by an NIR source that is not associated with the CO emission. Given only one line detection within the $\sim29$ GHz bandwidth, possible solutions will be 
$z_{\mathrm{CO(2-1)}}=1.512$ or $z_\mathrm{CO(3-2)}=2.767$, but these need to be verified by detecting other spectral lines in future observations \citep[e.g.,][]{2014ApJ...781L..39T,2021ApJ...917...94M}.

\begin{deluxetable}{ccccc}
\tablenum{3}
\caption{\label{table:table3}Results of Line Identification.}
\tablehead{\colhead{ID} &\colhead{CO transition}&\colhead{$z_\mathrm{CO}$}  &\colhead{$L'_{\mathrm{CO}(J_\mathrm{up}-J_\mathrm{low})}$}  &\colhead{$M_\mathrm{mol}$\tablenotemark{1}} \\
\colhead{} &\colhead{($J_\mathrm{up}-J_\mathrm{low}$)} &\colhead{} &\colhead{($10^{10}$ K km s$^{-1}$ pc$^2$)} &\colhead{($10^{10}M_\odot$)}}
\startdata
        1 &3-2 &2.124 &$1.45\pm0.19$&$12.5\pm1.6$\\
        2 &2-1 &1.113 &$0.62\pm0.06$&$2.9\pm0.3$\\
       4 &2-1 &1.146 &$1.31\pm0.26$&$6.2\pm1.2$\\
\enddata
\tablenotetext{1}{CO-based molecular gas mass (Section \ref{subsec:mgas}). Alternatively, galaxy 1 has  $M_\mathrm{mol}=(9.4\pm1.3)\times10^{10}M_\odot$ based on its [CI](1-0) line luminosity $L'_\mathrm{[CI](1-0)}=(4.04\pm0.55)\times10^9$ K km s$^{-1}$ pc$^2$.}
\end{deluxetable}

\par
The line luminosity is calculated from the velocity-integrated line flux $S\Delta v$, following \citet{2005ARA&A..43..677S}:
\begin{equation}\label{eq3}
\frac{L'}{\text{K km s}^{-1}\text{ pc}^2} = 
\frac{3.25\times10^7S\Delta v}{\text{Jy km s}^{-1}}
\frac{D_\mathrm{L}^2}{(1+z)^3\nu_\mathrm{obs}^2}\left[\frac{\mathrm{GHz}}{\mathrm{Mpc}}\right]^2,
\end{equation}
where $D_\mathrm{L}$ is the luminosity distance in megaparsecs corresponding to redshift $z$, and $\nu_\mathrm{obs}$ is the observed frequency of the line in gigahertz. We list the results of the line identification in Table \ref{table:table3}.  With the spectroscopic redshift determined from the CO line, we fix the redshift and refit the photometry for galaxies 1, 2, and 4, including band 3, 4, and 7 continuum measurements. In the following sections, we will focus on the derived physical properties of these three galaxies. 

\subsection{Stellar Mass and SFR from SED Modeling}\label{subsec:mstarsfr}
\begin{deluxetable}{cccccc}
\tablenum{4}
\caption{\label{table:table4}Derived Physical Properties from SED Fitting. } 
\tablehead{\colhead{ID} &\colhead{$M_\star$} &\colhead{SFR} &\colhead{$M_\mathrm{dust}$}&\colhead{$L_\mathrm{dust}$} &\colhead{$A_V$}\\
\colhead{}&\colhead{($10^{10}M_\odot$)}&\colhead{($M_\odot$ yr$^{-1}$)}&\colhead{($10^{8}M_\odot$)}&\colhead{($10^{11}L_\odot$)}&\colhead{(mag)}}
\startdata
        1 &$9.73^{+2.22}_{-2.47}$ &$126^{+37}_{-54}$ &$3.13^{+0.61}_{-0.54}$ &$12.79^{+2.90}_{-4.00}$&$2.39^{+0.16}_{-0.25}$\\
        2 &$4.60^{+1.39}_{-1.00}$&$22^{+12}_{-8}$&$2.35^{+0.54}_{-0.52}$&$2.97^{+1.36}_{-0.69}$&$2.15^{+0.29}_{-0.30}$\\
        4 &$21.98^{+3.92}_{-3.88}$&$0.7^{+0.8}_{-0.6}$ &$0.27^{+0.92}_{-0.01}$&$1.27^{+0.29}_{-0.27}$&$1.28^{+0.18}_{-0.20}$\\
        4.E &$0.03^{+0.02}_{-0.01}$&$9^{+1}_{-1}$ &$0.13^{+0.16}_{-0.05}$&$0.64^{+0.07}_{-0.06}$&$0.77^{+0.07}_{-0.05}$\\
        4.SW &$0.18^{+0.11}_{-0.08}$&$4^{+2}_{-2}$ &$0.07^{+0.11}_{-0.03}$&$0.33^{+0.20}_{-0.15}$&$0.56^{+0.30}_{-0.25}$\\
\enddata
\tablecomments{The SFRs are averaged over the past 30 Myr. For the E and SW clumps near galaxy 4, we also fix the redshifts at $z=1.146$.}
\end{deluxetable}

\begin{figure}
    \plotone{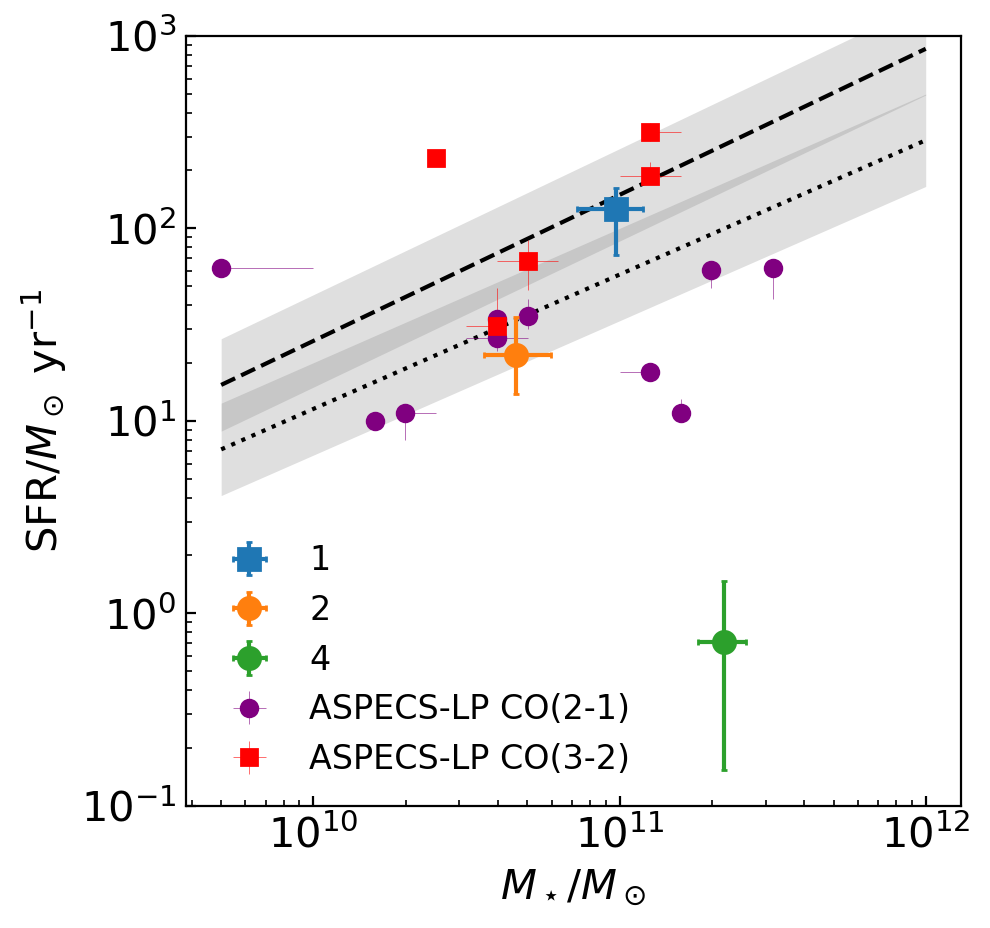}
    \caption{The positions of the three galaxies with CO redshift in the SFR-$M_\star$ plane. The $M_\star$ and SFRs are derived from SED modeling (Table. \ref{table:table4}). Circles mark galaxies detected with CO(2-1) while squares mark those with CO(3-2). The MS relations at $z=1.1$ and $z=2.1$ are shown with dotted and dashed lines, respectively. The shaded regions indicate the scatter of the MS.}
    \label{fig:fig6}
\end{figure}

\par
We show the SED fit and joint posterior distributions of the derived parameters in Figure \ref{fig:figa1} and \ref{fig:figa2}. The results from SED fitting are summarized in Table \ref{table:table4}. All reported values and errors represent median values and $68\%$ confidence intervals, respectively, from marginalized posterior PDFs.

\par
In Figure \ref{fig:fig6} we plot the three galaxies in the SFR-$M_\star$ plane, together with the MS relations at $z=1.1$ and $z=2.1$ from \citet{2014ApJS..214...15S}. For comparison, we also show the CO-selected galaxy sample from ASPECS-LP \citep{2019ApJ...882..140B} that surveyed a larger area at a similar depth. The ASPECS-LP $M_\star$ and SFR are derived using the \texttt{MAGPHYS} code \citep{2015ApJ...806..110D}. Different star formation history (SFH) and dust attenuation models can lead to a $\sim0.3$ dex systematic offset in the SED-derived $M_\star$ and SFR \citep{2019A&A...632A..79B}. Tests on simulated galaxies show the nonparametric SFH used in this study gives a larger $M_\star$ than simple exponential SFH models \citep[$\sim0.3$ dex;][]{2019ApJ...876....3L,2020ApJ...904...33L}. However, the results should still agree within error for the same object, as long as each method allows a wide range of SFHs and dust curves.

\par
All three galaxies with CO-based redshift are massive, with $M_\star>10^{10.5}M_\odot$ but different levels of star formation. Galaxies 1 and 2 lie within the scatter of the MS at their redshifts. In contrast, galaxy 4 is quiescent, with specific SFR $\mathrm{sSFR}=\mathrm{SFR}/M_\star=0.9_{-0.8}^{+2.8}\times10^{-12}$ yr$^{-1}$ and log($\Delta$MS)=log(SFR/SFR$_\mathrm{MS}$) = $-2.7^{+0.6}_{-0.9}$ (or sSFR$=6.6_{-1.5}^{+2.5}\times10^{-11}$ yr$^{-1}$ and log($\Delta$MS) = $-0.9\pm0.1$ when using UV+IR SFR estimates). Compared with the ASPECS-LP galaxies, galaxies 1 and 2 have comparable $M_\star$ and SFR, but galaxy 4 is a significant outlier, with SFR much lower than the median value (30 $M_\odot$ yr$^{-1}$) for galaxies detected with CO(2-1) in the ASPECS-LP survey.

\begin{figure}
    \plotone{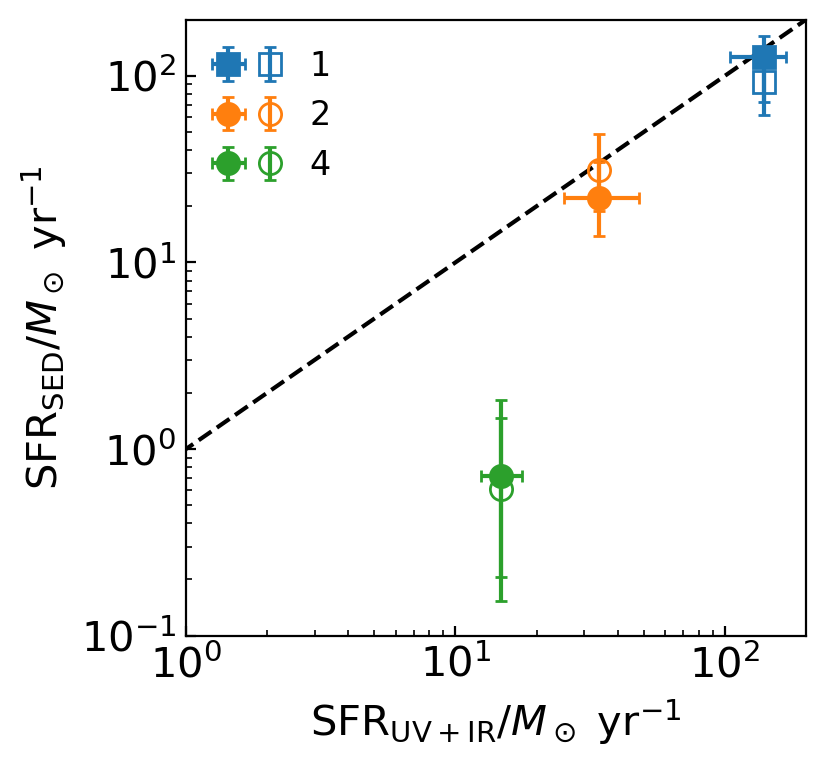}
    \caption{SFR$_\mathrm{SED}$ vs. SFR$_\mathrm{UV+IR}$. The dashed line denotes 1:1 equality. The filled and void symbols show  SFR$_\mathrm{SED}$ averaged over the past 30 Myr and 100 Myr, respectively}
    \label{fig:fig7}
\end{figure}

\par 
We further compare our SED-based SFR with those derived from UV+IR conversions in Figure \ref{fig:fig7}. The UV+IR SFR is calculated following \citet{2005ApJ...625...23B}:
\begin{equation}\label{eq4}
    \frac{\mathrm{SFR}_\mathrm{UV+IR}}{M_\odot \text{ yr}^{-1}}=\frac{1.09\times10^{-10}(2.2L_\mathrm{UV}+L_{\mathrm{IR}}
    )}{L_\odot},
\end{equation}
where $L_\mathrm{UV}$ is the rest-frame luminosity integrated in the rest-frame wavelength range 121.6-300 nm without dust correction and $L_\mathrm{IR}$ is the luminosity integrated { in the rest-frame wavelength range 8-1000 $\mu$m of posterior SED models}. Galaxies 1 and 2 have SFR$_\mathrm{SED}/$SFR$_\mathrm{UV+IR}$ of $91_{-24}^{+8}\%$ and $67_{-16}^{+9}\%$, respectively. Comparatively, galaxy 4's SFR$_\mathrm{SED}$ is only $5_{-4}^{+5}\%$ of SFR$_\mathrm{UV+IR}$. This trend decreases while remaining visible if we change the average interval of SFR$_\mathrm{SED}$ from 30 Myr to 100 Myr. The difference between the two SFR methods increases with decreasing sSFR, because of the contribution of dust heating from old stars \citep{2014MNRAS.445.1598H,2019ApJ...877..140L}, and this behavior is also found when using the \texttt{MAGPHYS} code \citep{2019ApJ...882...65M}. Since the dust attenuations of old and young stars are modeled separately and both low and high attenuation values are allowed for the two stellar populations, the error range of SFR$_\mathrm{SED}$ serves as a conservative estimate of the SFR levels.  Therefore, we continue to choose SFR$_\mathrm{SED}$ as our SFR indicator. 

\subsection{Molecular Gas Mass}\label{subsec:mgas}
In order to derive their molecular gas mass, we first convert the CO luminosity to the ground transition, assuming line ratios $r_{21}=0.76\pm0.09$ and $r_{31}=0.42\pm0.07$ appropriate for high-redshift normal SFGs \citep{2015A&A...577A..46D,2016ApJ...833...70D}. The total molecular gas mass $M_\mathrm{mol}$ is calculated using the conversion factor of normal SFGs: $\alpha_{\mathrm{CO}}=M_\mathrm{mol}/L'_{\mathrm{CO(1-0)}}=$ 3.6 $M_\odot/(\mathrm{K}$ km s$^{-1}$ pc$^2)$; \citep{2010ApJ...713..686D}. The resulting gas masses are listed in Table. \ref{table:table2}. For galaxy 1, we can also estimate $M_\mathrm{mol}$ from the [CI] line flux. The conversion factor $\alpha_{\mathrm{[CI](1-0)}}=M_\mathrm{mol}/L'_{\mathrm{[CI](1-0)}}$ is calculated using the equation 11 of \citet{2022MNRAS.517..962D}:
\begin{equation}\label{eq5}
 \frac{\alpha_{\mathrm{[CI](1-0)}}}{M_\odot(\text{K km s}^{-1} \text{pc}^2)^{-1}}
 =16.8\left(\frac{X_{\mathrm{CI}}}{1.6\times10^{-5}}\right)^{-1}\left(\frac{Q_{10}}{0.48}\right)^{-1},
\end{equation}
where $X_\mathrm{CI}$ is the average abundance ratio of atomic carbon and $Q_{10}$ is the excitation term. We adopt $X_\mathrm{CI}=10^{-4.8}$ and $Q_{10}=0.35$ following \citet{2021ApJ...909..181L}.
Using the band 4 data, we find the velocity-integrated [CI](1-0) line flux $S_\mathrm{[CI](1-0)}\Delta v=0.320\pm0.044$ Jy km s$^{-1}$. This translates to a [CI](1-0)-based molecular gas mass of $(9.4\pm1.3)\times10^{10}M_\odot$, roughly consistent with the CO-based gas mass.

\par
We note that molecular gas mass estimates suffer from substantial uncertainties in $\alpha_\mathrm{CO}$ and excitation.  In this study, we have adopted $\alpha_\mathrm{CO}$ and line ratios of normal SFGs. Some recent ALMA studies have revealed compact star formation with intense starburst-like conditions in high-redshift MS galaxies, or ``hidden starburst in the MS'' \citep[e.g., ][]{2021MNRAS.508.5217P}. In this case, starburst-like conversions should be used to derive gas mass, despite their SFR being within the scatter of the MS. If we instead assume SMG-like $\alpha_\mathrm{CO}=1.36$ $M_\odot/({\mathrm{K}}$ km s$^{-1}$ pc$^2)$, $r_{21}=0.84\pm0.13$ and $r_{31}=0.52\pm0.09$ \citep{2013MNRAS.429.3047B}, the gas mass will decrease by 2.96 and 3.28 times for CO(2-1) and CO(3-2), respectively. For the [CI](1-0) line, with SMG-like $X_\mathrm{[CI]}=10^{-4.2}$ and $Q_{10}=0.48$ \citep{2011ApJ...730...18W,2018ApJ...869...27V} the estimated molecular gas mass of galaxy 1 becomes $(1.7\pm0.2)\times10^{10}M_\odot$, or 45\% of the CO-based gas mass.

\par

Our observations provide three tracers of the molecular gas, namely CO, [CI], and Rayleigh-Jeans tail dust continuum. However, there is a discrepancy between the derived $M_\mathrm{mol}$ using CO and dust in one of the two galaxies with continuum detection. Under normal SFG conversions of CO, galaxy 2 has a gas-to-dust mass ratio $\delta_\mathrm{GDR}=126^{+39}_{-26}$ with $M_\mathrm{dust}$ from the SED fitting, which is consistent with $\delta_\mathrm{GDR}=103$ from scaling relations \citep[e.g., ][]{2008A&A...488..463M,2011ApJ...737...12L,2015ApJ...800...20G}. In contrast, galaxy 1 has $\delta_\mathrm{GDR}=396^{+105}_{-81}$, more than three times higher than $\delta_\mathrm{GDR}=111$ expected by the  $M_\star$-metallicity-redshift and metallicity-$\delta_\mathrm{GDR}$ scaling relations.
\par For galaxy 1, SMG-like conversions bring $\delta_\mathrm{GDR}=121^{+32}_{-25}$ into an agreement with the scaling relations, but this cannot be used to justify SMG-like conversion for galaxy 1. The dust-based $M_\mathrm{mol}$ is the least robust here, because of low S/N ($<2$) in all Herschel bands and the uncertainty in the dust SED fitting \citep[e.g.,][]{2016A&A...587A..73B}. The position of galaxy 1 with respect to the $L_\mathrm{dust}-L^{'}_{\mathrm{CO(1-0)}}$ relation also disfavors a starburst nature of galaxy 1.  \citet{2010MNRAS.407.2091G} find $\log(L_\mathrm{dust})=1.15\log(L^{'}_{\mathrm{CO(1-0)}})+0.02\pm1.1$ for normal SFGs and the same slope, but with an intercept of $0.63\pm0.12$, for starbursts. For galaxy 1's $L_\mathrm{dust}=12.11^{+3.57}_{-3.55}\times10^{11}L_\odot$ and $L^{'}_{\mathrm{CO(3-2)}}=(1.45\pm0.19)\times10^{10}$ K km s$^{-1}$ pc$^2$, we derive $\log(L_\mathrm{dust}/L^{'}_{\mathrm{CO(1-0)}}{}^{1.15})=-0.02^{+0.12}_{-0.16}$ under normal SFG conversion or $0.08^{+0.12}_{-0.17}$ under SMG conversion, thus the galaxy seems to be closer to a normal SFG. In the following discussion, we continue using $M_\mathrm{mol}$ from normal SFG conversions.

\section{Discussion}\label{sec:discussion}

\begin{figure*}
    \plotone{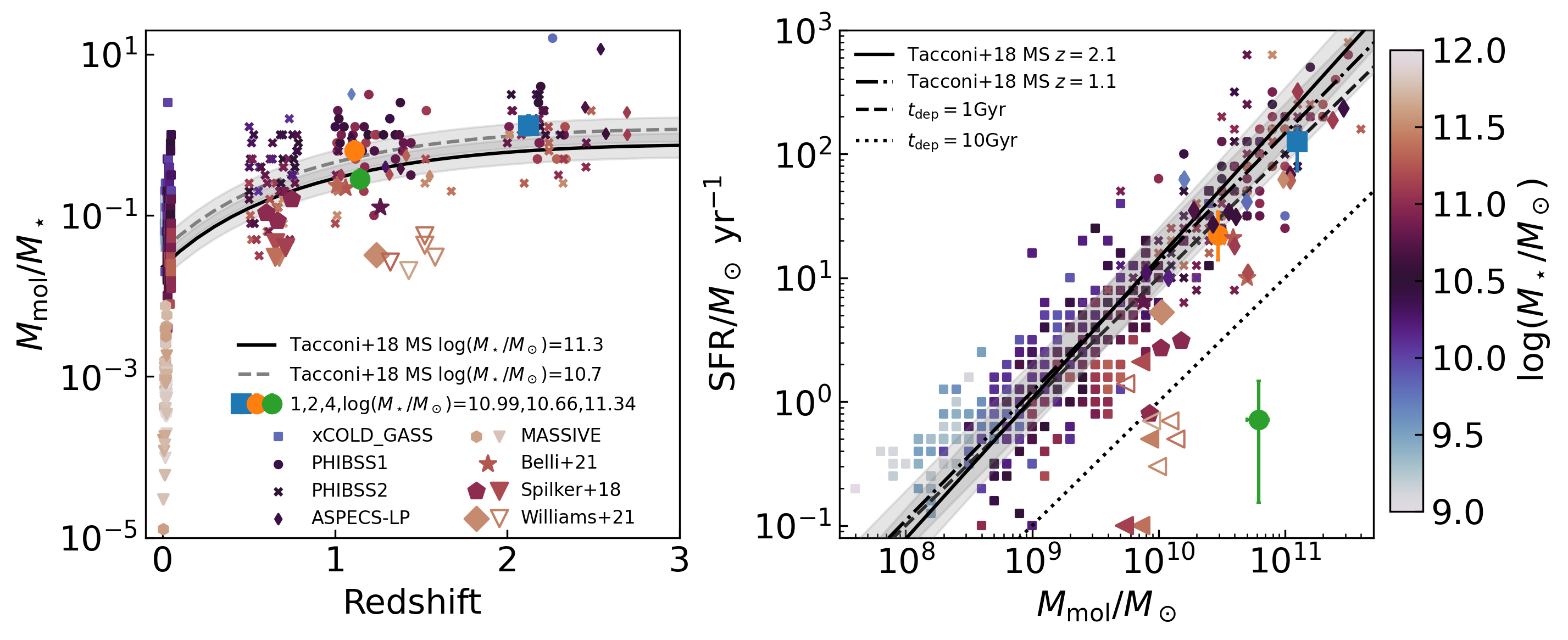}
    \caption{Left: molecular gas fraction as a function of redshift. For comparison, we show the scaling relation and its scatter from \citet{2018ApJ...853..179T}, local SFGs \citep[xCOLD GASS; ][]{2017ApJS..233...22S}, intermediate- and high redshift star-forming galaxies \citep[PHIBSS1/2; ][]{2013ApJ...768...74T,2019A&A...622A.105F}, local QGs \citep[MASSIVE; ][]{2019MNRAS.486.1404D}, intermediate- and high-redshift QGs \citep{2018ApJ...860..103S,2021ApJ...909L..11B,2021ApJ...908...54W}. The triangles indicate upper limits. Right: SFR as a function of $M_\mathrm{mol}$. The symbols are the same as in the left panel.}
    \label{fig:fig8}
\end{figure*}

\subsection{Comparison with Gas Scaling Relations}\label{subsec:gasscaling}
The left panel of Figure \ref{fig:fig8} shows the gas ratio $\mu_\mathrm{mol}=M_\mathrm{mol}/M_\star$ of our sample and values of SFGs and QGs across cosmic time, color-coded by $\log(M_\star/M_\odot)$. The SFG sample is collected from \citet{2018ApJ...853..179T}, which uses the \citet{2015ApJ...800...20G} metallicity-dependent prescription---$\alpha_\mathrm{CO}=3.8-12.4$ $M_\odot$(K km s$^{-1}$ pc$^{2}$)$^{-1}$ for $z=0-4$ and $\log(M_\star/M_\odot)=9-12$. The $M_\mathrm{mol}$ of QGs in the comparison sample is derived using Galactic $\alpha_\mathrm{CO}$. At $z=1-1.5$, the SFR$>30$ $M_\odot$ yr$^{-1}$ and $M_\star>2.5\times10^{10}M_\odot$ cuts of the PHIBSS1 survey make their preselection slightly higher than MS SFR at this redshift,  resulting in a median $\mu_\mathrm{mol}$ of 0.79. This is higher than $\mu_\mathrm{mol}=0.63_{-0.15}^{+0.20}$ of galaxy 2 or $0.28_{-0.06}^{+0.09}$ of galaxy 4. Similar differences have been found by the ASPECS-Pilot survey \citep{2016ApJ...833...70D}, which found that CO(2-1) emitters from their ALMA band 3 line search in the HUDF field have a median SFR of $34$ $M_\odot$ yr$^{-1}$ and a $\sim$two times lower median gas ratio than PHIBSS1. Also, in the subsequent larger ASPECS-LP sample \citep{2019ApJ...882..136A}, the galaxies detected with CO(2-1) have median $z=1.20$ and $\mu_\mathrm{mol}=0.48$. At $z>2$, the median gas ratio of PHIBSS1 galaxies is 1.26, comparable with $\mu_\mathrm{mol}=1.28_{-0.29}^{+0.49}$ for galaxy 1, as at this redshift their SFR and $M_\star$ cuts do not bias the sample to galaxies above the MS. Our sample further contains one galaxy that is significantly below the MS, showing that a blind line search is able to cover a wider range of star formation levels.

\par
The right panel of Figure \ref{fig:fig8} shows SFR as a function of $M_\mathrm{mol}$. The two quantities are closely correlated with the modest time evolution of depletion time $t_\mathrm{dep}=M_\mathrm{mol}/\mathrm{SFR}\propto(1+z)^{-0.63}$ \citep{2018ApJ...853..179T}. In our sample, galaxies 1 and 2 show a lower SFR at a given $M_\mathrm{mol}$ compared to the targeted SFGs in the PHIBSS1 survey, but their $t_\mathrm{dep}=1.0_{-0.3}^{+0.7}$ and $1.3_{-0.5}^{+0.8}$ Gyr agree with the scaling relation, which expects $\sim0.7$ and $\sim1$ Gyr, respectively. This might be caused by the difference in the methods for deriving SFR and $M_\star$. As mentioned in Section \ref{subsec:mstarsfr}, the full UV-to-FIR SED fitting method with nonparametric SFH used in this study typically gives lower SFR and larger $M_\star$. Such a trend is also found by ASPECS-LP \citep{2019ApJ...882..136A}, as they use \texttt{MAGPHYS}, which also includes dedicated treatments of SFH and dust attenuation. { In summary, while the target selection and methodology of analysis are different}, our sample shows that the current scaling relations successfully describe the evolution of gas content  as a function of redshift, mass, and SFR in SFGs.
{
\subsection{On the Nature of Galaxy 4}\label{subsec:gal4} 
\subsubsection{Relation between Galaxy 4 and the Two Neighboring Sources}\label{subsubsec:gal4clump}
The most surprising finding from the analysis above is the large gas reservoir in the massive galaxy 4 with low sSFR. The analysis is based on the assumption that the two blue clumps (E and SW; Figure \ref{fig:fig4}) are not physically associated with the CO(2-1) emission. If we ignore the clustering effect, so that galaxies are uniformly distributed, the probability of the chance alignment of a galaxy $i$ can be expressed as \citep{2002AJ....123.1111B}: \begin{equation}
    P_\mathrm{chance}=1-\exp(-\sigma_{i}\pi r_{i}^{2}),
\end{equation}where $\sigma_i$ is the surface density of sources brighter than galaxy $i$ and $r_i$ is the sky separation between the CO detection and galaxy $i$. Using the $B$-band image we measure $\sigma_i=0.011$ arcsec$^{-2}$, $r_{i}=1\farcs34$ for the E clump and $\sigma_i=0.006$ arcsec$^{-2}$, $r_{i}=1\farcs03$ for the SW clump. The random coincidence rate of two such galaxies is $1.18\times10^{-3}$. This low probability rejects the null hypothesis that the two clumps are randomly aligned at the $3.2\sigma$ level. To assess whether this is a rare case of random alignment, we further examine the SED and physical properties of  the two blue clumps and the central massive galaxy.
\begin{figure*}
    \plotone{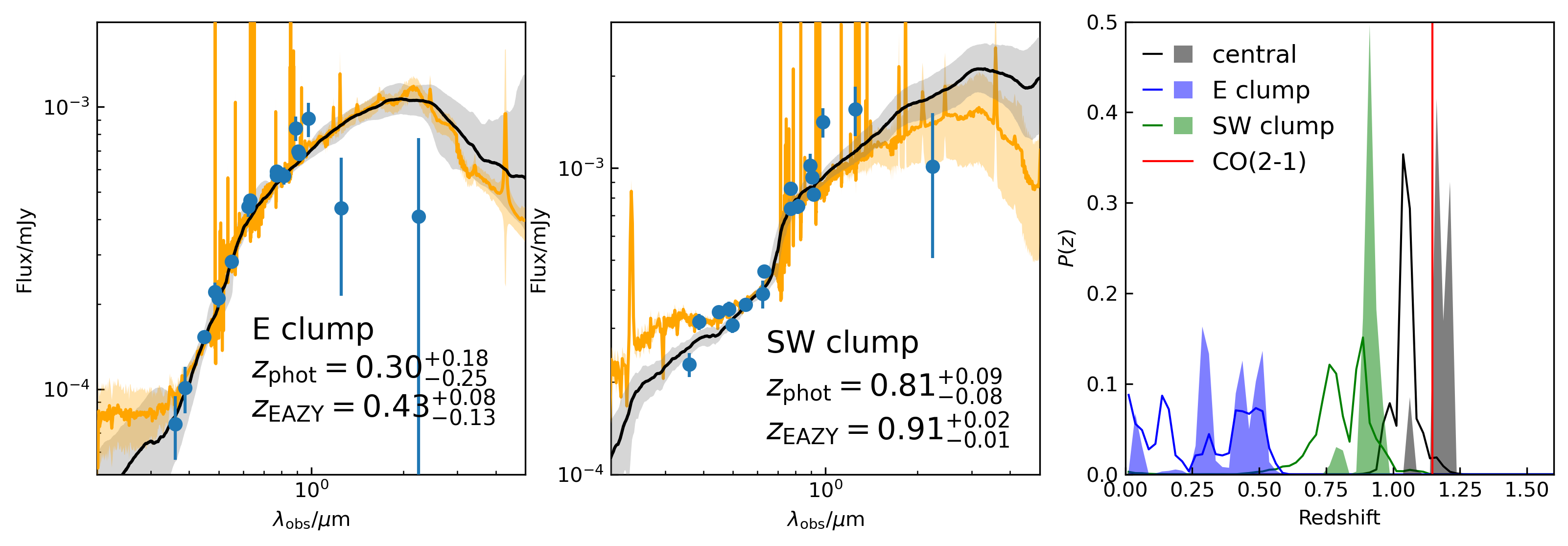}
    \caption{ Left: deblended SED for the E clump. The median (or best-fit for \texttt{EAZY}) and 16th-84th percentile ranges of the SED fit are shown as the solid lines and shaded regions, with black color for the method in section \ref{subsec:SEDfitting} and orange color for the \texttt{EAZY} code. Middle: the same as left but for the SW clump. Right: the photometric redshift PDF $P(z)$ of the central galaxy, E, and SW clumps. The solid lines show the photometric redshift from the SED fitting method in section \ref{subsec:SEDfitting} and the shaded regions are the results from the \texttt{EAZY} code.}
    \label{fig:fig9}
\end{figure*}
\begin{figure}
    \plotone{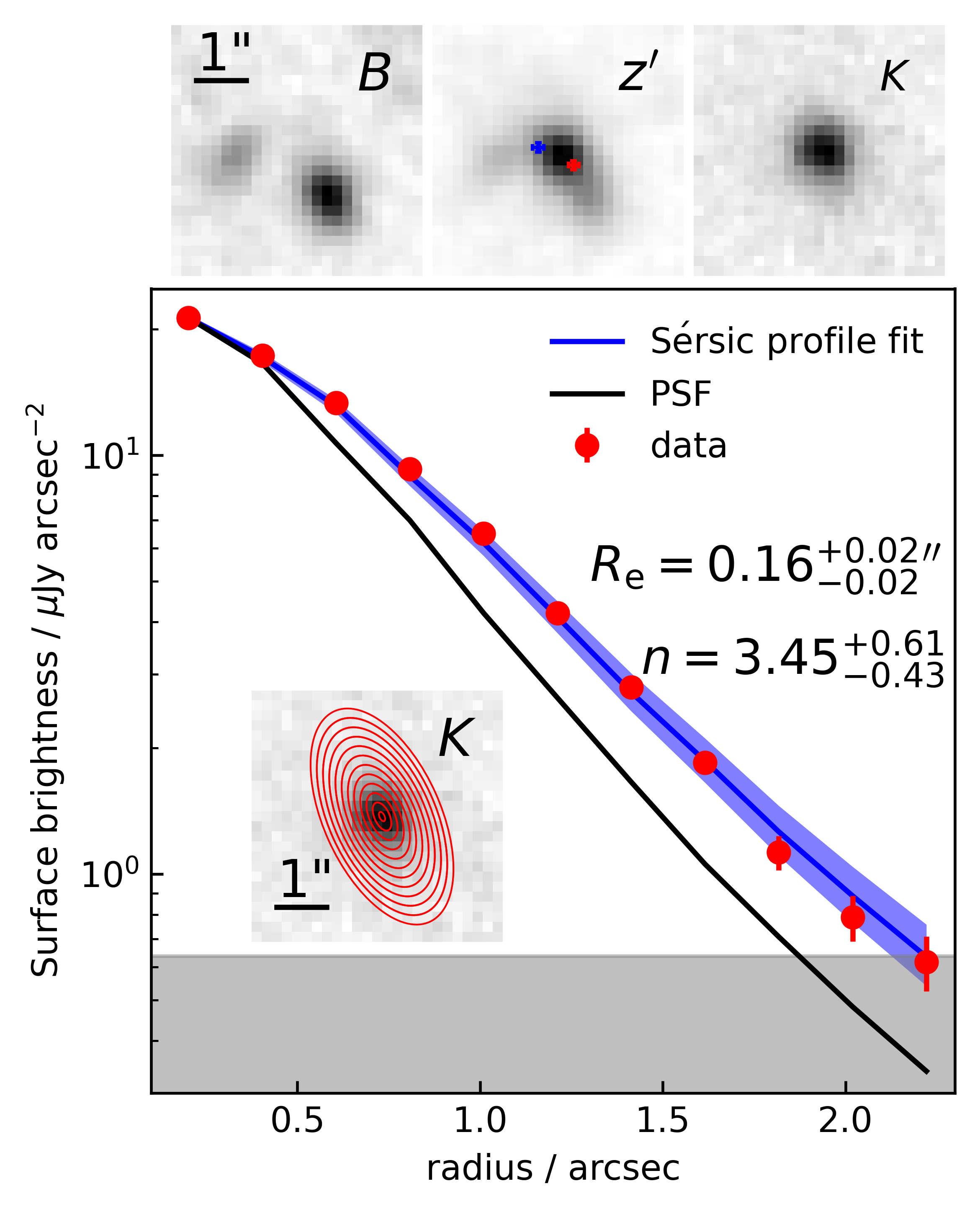}
    \caption{ Top: $5\arcsec\times5\arcsec$ cutout stamps centered at the CO position in $B$, $z'$ and $K$ bands. The blue and red error bars in the $z'$ band image mark the centroids of the two velocity components of the CO emission. Bottom: $K$-band surface brightness distribution along the major axis. The data points are derived from the $K$-band image without neighbor subtraction using the elliptical annuli shown in the insert panel. The black line indicates the PSF. The blue solid line and shaded region are the median and 16th-84th percentile range of the 2D S\'{e}rsic model of the central galaxy, which is derived by fitting three objects jointly (section \ref{subsubsec:optnir}). The gray region represents the RMS of a single pixel. }
    \label{fig:fig10}
\end{figure}
\par As shown in the right panel of Figure \ref{fig:fig9}, the photometric redshifts from our SED fitting method are $0.30_{-0.25}^{+0.18}$ and $0.81_{-0.08}^{+0.09}$ for the E and SW clumps, respectively. Only the SW clump has a nonzero probability at $z=1.146$. Meanwhile, the photometric redshifts from \texttt{EAZY} do not overlap with the redshift of the CO(2-1) emission line. On the other hand, by fitting the deblended photometry of the two clumps at fixed $z=1.146$, we derive SFR=$9^{+1}_{-1}M_\odot$ yr$^{-1}$, $M_\star=0.03^{+0.02}_{-0.01}\times10^{10}M_\odot$ for the E clump and SFR=$4^{+2}_{-2}M_\odot$ yr$^{-1}$, $M_\star=0.18^{+0.11}_{-0.08}\times10^{10}M_\odot$ for the SW clump (Table \ref{table:table4}). 
\par
High-redshift galaxies have been shown to host blue clumpy structures, which are the sites of star formation \citep[e.g.,][]{2009ApJ...701..306E}. However, the properties of the two clumps surrounding galaxy 4 are incompatible with such a picture. Firstly, the central galaxy appears like a spheroid (Figure \ref{fig:fig10}) without a disk component, which is the expected location of star-forming clumps. Also, the clumps are outside the central galaxy, because the separations of $\sim1\arcsec$ are larger than the $R_\mathrm{e}=0\farcs16-0\farcs35$ of individual components. Secondly, the sizes of the two clumps are too large compared to the central galaxy. For the central spheroid, we measure $R_\mathrm{e}=0.16^{+0.02}_{-0.02}$ arcsec, while the E clump has $R_\mathrm{e}=0.30^{+0.09}_{-0.08}$ arcsec and the SW clump has $R_\mathrm{e}=0.35^{+0.01}_{-0.01}$ arcsec. Assuming the spherical symmetry of the central spheroid and thickness$=0.1R_\mathrm{e}$ disk of the two clumps, and using $M_\star$ from the SED fitting (Table \ref{table:table4}), the Roche limit is $\sim2\farcs8$ for the E clump and $\sim1\farcs8$ for the SW clump in the image plane. This means the tidal force from the central massive galaxy will destruct these two clumps to form a ring or disk if the two clumps are located inside the sphere with the Roche radius. However, such features are not seen in the image. Last, the centroids of the blue and red components of the CO(2-1) emission are closer to the central galaxy (the $z'$ band cutout in Figure \ref{fig:fig9}). If the CO emission is from the two clumps, the $\mu_\mathrm{mol}$ will be $\sim100$ for the E clump and $\sim15$ for the SW clump under $\alpha_\mathrm{CO}=3.6$ $M_\odot$(K km s$^{-1}$ pc$^{2}$)$^{-1}$ and $r_{21}=0.76$. Such an extreme molecular gas-rich dwarf has not been detected in the local universe \citep{2017ApJS..233...22S}.
\par
Overall, the positions and properties of the two clumps near galaxy 4 suggest that they are likely to be interlopers rather than companions or parts of the central massive galaxy, from which the CO emission originates. Future optical/NIR spectroscopy is needed to obtain accurate redshifts and verify the discussion presented above.
}
\subsubsection{Large Gas Reservoir in a Massive Galaxy below the MS?}\label{subsubsec:gal4}
CO/dust observations of QGs outside the local universe based on preselection of sSFR typically find a low gas fraction of $\mu_\mathrm{gas}\lesssim10\%$ \citep[][ see also the left panel of Figure \ref{fig:fig8} for CO samples]{2015ApJ...806L..20S,2018ApJ...860..103S,2019ApJ...873L..19B,2021ApJ...910L...7C,2021ApJ...908...54W}. On the other hand, higher values of $\mu_\mathrm{mol}$ in QGs have also been reported \citep{2017ApJ...849...27R,2018NatAs...2..239G,2018ApJ...856..118H,2021ApJ...909L..11B}. Here, galaxy 4 has $M_\mathrm{mol}=(6.2\pm1.2)\times10^{10}$ $M_\odot$ and $\mu_\mathrm{gas}$ of $0.28_{-0.06}^{+0.09}(\alpha_\mathrm{CO}/3.6$ $M_\odot$ $({\mathrm{K}}$ km s$^{-1}$ pc$^2)^{-1})((0.76\pm0.09)/r_{21})$, adding one more candidate to gas-rich QGs. With the results, we infer a very low star formation efficiency (SFE) of galaxy 4 with SFE$=1/t_\mathrm{dep}=1.15^{+1.44}_{-0.89}\times10^{-11}$ yr$^{-1}$. The SFE is not only lower than the targeted SFGs in the PHIBSS surveys,  but also lower than any blindly detected galaxies from ASPECS-LP, which has the lowest SFE$=2.2^{+0.5}_{-0.3}\times10^{-10}$ yr$^{-1}$. In contrast, preselected QGs detected in CO at $z\gtrsim1$ show SFEs closer to normal SFGs (the right panel of Figure \ref{fig:fig8})

\par
If the gas reservoir is left over after the end of the main star formation episode without being depleted or destroyed, there must be some mechanisms to halt new star formation. In bulge-dominated systems, the gas disk can be stabilized against collapse \citep{2009ApJ...707..250M,2014ApJ...785...75G}, so further star formation is dynamically suppressed. However, the gas fraction of galaxy 4 is too high for morphological quenching to effectively reduce the star formation \citep[$\lesssim5\%$; ][]{2013MNRAS.432.1914M,2020MNRAS.495..199G}, though its contribution to maintaining quiescence is possible \citep{2021MNRAS.500.2000G}. From the SFH (Figure \ref{fig:figa2}), we see that the bulk of stars were formed $\gtrsim2$ Gyr ago and the SFR drops after $z\sim2$, with a mass-weighted stellar age of $3.3^{+1.4}_{-0.4}$ Gyr. Other sources such as supernovae or active galactic nuclei \citep{2010A&A...521A..65N} can also inject kinematic energy into the ISM and prevent collapse, but these processes take place on shorter timescales \citep[$\lesssim100$ Myr; ][]{2015A&A...574A..32G} and may also remove the cold gas, so they are at least not the main reason for galaxy 4's quiescence while maintaining a high gas fraction. Finally, the $\gtrsim1$ Gyr lookback time is much longer than the remaining lifetime of the gas reservoir after the quenching inferred from post-starburst galaxies \citep[$\sim150$ Myr; ][]{2022ApJ...925..153B}. Thus, the gas reservoir in galaxy 4 is no likely to be the remnant of the past major star formation event, but the listed mechanisms might have played a role in maintaining the low SFE.

\par
Alternatively, the gas reservoir might be acquired in late times \citep{2022ApJ...940...39W}. Since galaxy 4 is already very massive ($M_\star>10^{11}M_\odot$), its halo mass will be $M_\mathrm{halo}>10^{12.5}M_\odot$ assuming a stellar-to-halo mass ratio of -1.5 dex \citep{2013ApJ...770...57B}. At this $M_\mathrm{halo}$, shock heating can suppress further cold gas supply via accretion from the surrounding environment \citep{2005MNRAS.363....2K,2006MNRAS.368....2D}. Rather than accretion, several minor mergers could have added gas to galaxy 4. According to the MS gas scaling relation, a normal galaxy with one-houndredth to one-quarter of galaxy 4's $M_\star$ and redshift $1.146-2$ carries $M_\mathrm{gas}\sim0.3\times10^{10}$ $M_\odot-4.8\times10^{10}$ $M_\odot$. Integrating the merger rate found by the Illustris simulation \citep{2015MNRAS.449...49R} over this mass and redshift range, there are 2.9 mergers expected and $3.4\times10^{10}$ $M_\odot$ of added molecular gas, which is only half of the gas mass of galaxy 4. The minor merger scenario also struggles to explain the low SFR, though it does not necessarily elevate the SFR immediately, due to dynamical effects \citep{2015MNRAS.449.3503D,2018MNRAS.476..122V}. If minor mergers really happened to galaxy 4, there should be some imprints left in it, such as stellar gas misalignment \citep[e.g.,][]{2021ApJS..254...27K}. The CO data show that a rotating gas disk might exist in galaxy 4 (Figure \ref{fig:fig3}), but currently we cannot perform kinematic diagnostics, due to a lack of optical/NIR spectroscopic data. We conclude that none of the mechanisms discussed above can explain galaxy 4's high gas fraction alone, and future high-resolution observations are needed to reveal the nature of galaxy 4. Such observations may include high-resolution optical/NIR imaging to study stellar morphology; confirmations of quiescence and relations with neighbor sources using optical/NIR spectroscopy;  and interferometric observation of CO line kinematics, especially for diagnostics of stability, temperature, and the density of its gas disk.  

\subsubsection{Caveats of Galaxy 4's SFR Estimate}\label{subsec:caveat}
\par
We caution that the current identification of galaxy 4 as a QG is only tentative.  The assertion of the low SFR is based on very faint rest-frame UV fluxes and the obscured SFR of galaxy 4 is left not well constrained. In this $M_\star$ range, almost all of the star formation is obscured \citep[$>90\%$; e.g., ][]{2017ApJ...850..208W}, so FIR/radio observations are needed to measure the total SFR. The galaxy is not detected in Herschel SPIRE bands, which have a $1\sigma$ depth of $\sim7$ mJy. The $3\sigma$ upper limits of ALMA band 3/4 continuum fluxes are still one to two orders of magnitudes higher than that predicted by SED fitting (Figure \ref{fig:figa1}). Assuming a modified blackbody model with dust temperature $T_\mathrm{dust}=20-60$ K and emissivity index $\beta=1.5-2.0$, the $3\sigma$ upper limit of the 2 mm continuum flux density $S_\mathrm{2mm}<0.1$ mJy translates to $L_\mathrm{dust}<0.4-26.9\times10^{11}L_\odot$ or SFR$<4-294M_\odot$ yr$^{-1}$ for dust heated by star formation. 
\par
A similar constraint on SFR can be obtained from JVLA 3 GHz observations \citep{2017ApJ...850..178A} of the SSA22 field. The radio observations have a PB response of $\sim0.13$ at the position of galaxy 4, resulting in a $3\sigma$ upper limit of 33 $\mu$Jy at 3 GHz. Assuming a radio spectral index of -0.7 (i.e., $S_\nu\propto\nu^{-0.7}$) and SFR/($M_\odot$ yr$^{-1}$)=$6.35\times10^{-22}L_\mathrm{1.4GHz}$/(W Hz$^{-1}$) \citep{2011ApJ...737...67M}, the flux upper limit implies an SFR upper limit of $\sim220$ $M_\odot$ yr$^{-1}$. Obviously, current FIR and radio data are not deep enough to constrain galaxy 4's SFR. Future follow-up observation at submillimeter/millimeter wavelengths will be critical to confirm its quiescence. If our current SFR estimate of galaxy 4 is broadly correct, this object will be particularly interesting for future studies, to test galaxy quenching scenarios, as discussed in the previous subsection.

\section{Summary}\label{sec:summary}
In this work, we have analyzed the data from $\sim29$ GHz bandwidth ALMA band 3 spectral scan observations toward SMG SSA22-AzTEC26 and detected four line emitters with S/N above 5.2 in a spectral cube smoothed with a 300 km s$^{-1}$ FWHM Gaussian filter, after masking the central SMG. We combine ALMA band 4/7 and multiwavelength ancillary data to derive the photometric redshift and their physical properties. Our main findings are:

\par 1. Using the photometric redshift, we identify that two of the sources are CO$(2-1)$ at $z=1.113$ and $z=1.146$ respectively. Another is CO(3-2) at $z=2.124$ and confirmed with [CI](1-0) emission line detected in the ALMA band 4 observation. The remaining source might be $z_{\mathrm{CO(2-1)}}=1.512$ or $z_{\mathrm{CO(3-2)}}=2.767$ and need to be verified by future observations.

\par
2. All three sources with a CO redshift solution are massive galaxies ($M_\star>10^{10.5}M_\odot$), with SFR in the order of $\sim1-100$ $M_\odot$ yr$^{-1}$. Two of them lie within the scatter of the MS and the most massive galaxy 4 is significantly below the MS. 

\par
3. Using the CO data, we estimate a molecular gas mass of $2.9-12.5\times10^{10}M_\odot$ under the assumption of normal SFG excitation and $\alpha_{\mathrm{CO}}$. We compare different gas tracers and conversion factors, finding that our current choice of normal SFG excitation and $\alpha_{\mathrm{CO}}$ seems to be appropriate. The gas fraction and $t_\mathrm{dep}$ of galaxies 1 and 2 are consistent with the gas scaling relation. However, the most massive one, galaxy 4, is likely quiescent but maintains a gas ratio of $\sim28\%$, comparable with SFGs on the MS.

\par
It is difficult to explain the large gas reservoir of galaxy 4 as being left over after quenching because of its high gas fraction and old stellar age. We speculate that the gas content was obtained in later times, via accretion or minor mergers, while various quenching mechanisms might have acted to suppress the star formation. Future high-resolution observations are needed to investigate the stellar and gas kinematics to understand the origin of the large gas reservoir and the exact quenching channels in this galaxy.

\begin{acknowledgements}
We thank the anonymous referee for the detailed and constructive comments that improved the quality of this paper. H.U. acknowledges the support from JSPS KAKENHI grant No. 20H01953. K.K. and R.K.  acknowledge the support by JSPS KAKENHI grant No. JP17H06130. Y.T. acknowledges the support from NAOJ ALMA Scientific Research grant No. 2018-09B and JSPS KAKENHI Nos. 17H06130, 19H01931, and 22H04939. This paper makes use of the following ALMA data: ADS/JAO.ALMA\#2019.1.01102.S and \#2021.1.01207.S. ALMA is a partnership of ESO (representing its member states), NSF (USA), and NINS (Japan), together with NRC (Canada), MOST and ASIAA (Taiwan), and KASI (Republic of Korea), in cooperation with the Republic of Chile. The Joint ALMA Observatory is operated by ESO, AUI/NRAO, and NAOJ. Herschel is an ESA space observatory with science instruments provided by European-led Principal Investigator consortia and with important participation from NASA. The Cosmic Dawn Center (DAWN) is funded by the Danish National Research Foundation under grant No. 140.
\end{acknowledgements}

\software{numpy \citep{harris2020array},
scipy \citep{2020SciPy-NMeth},
matplotlib \citep{Hunter:2007},
astropy \citep{2018AJ....156..123A}, photutils \citep{2022zndo...6825092B}, dynesty \citep{2020MNRAS.493.3132S}, SExtractor \citep{1996A&AS..117..393B}, PSFEx \citep{2011ASPC..442..435B}}

\begin{figure*}[h]
\figurenum{A1}
    \gridline{\fig{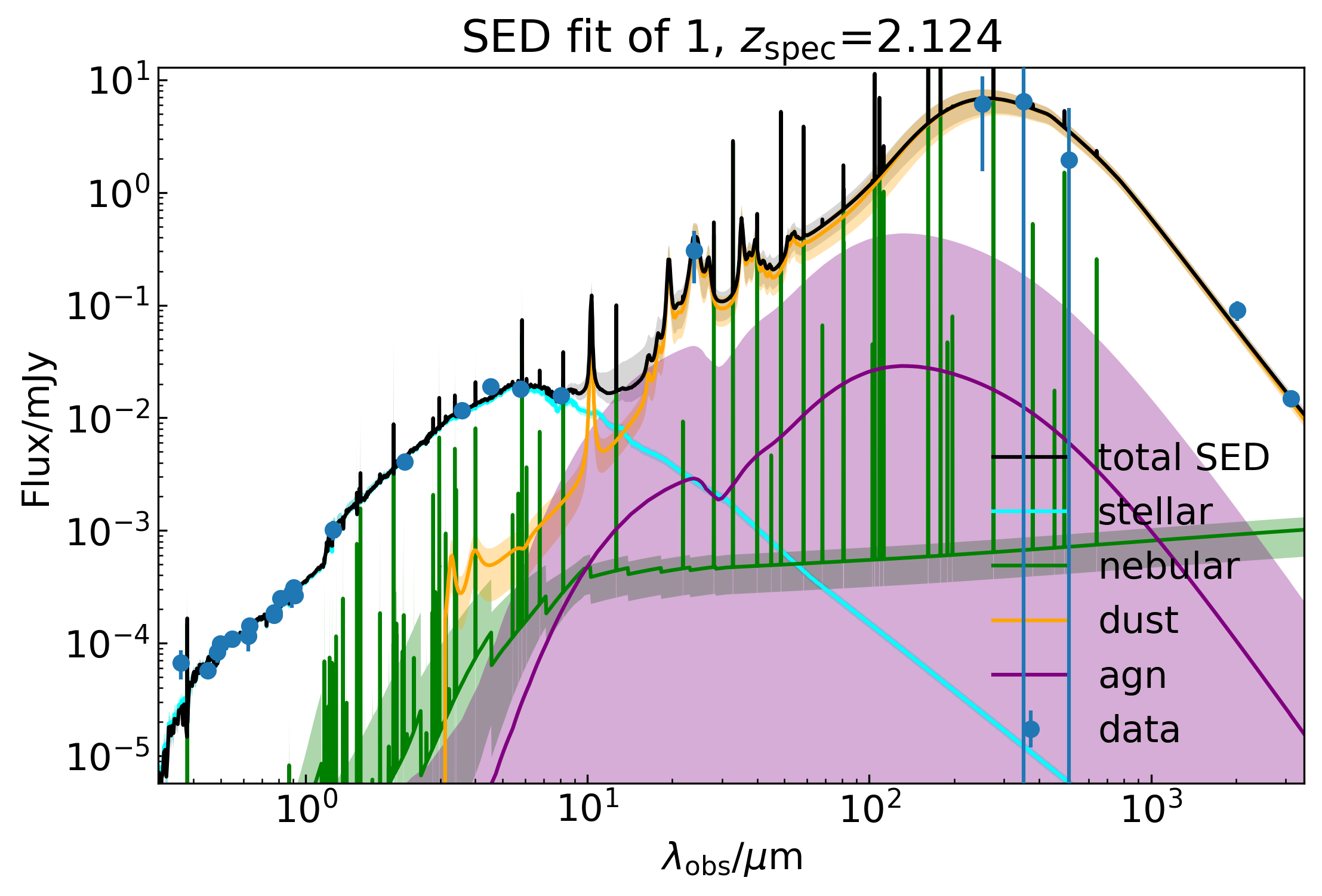}{0.5\textwidth}{(1)}\fig{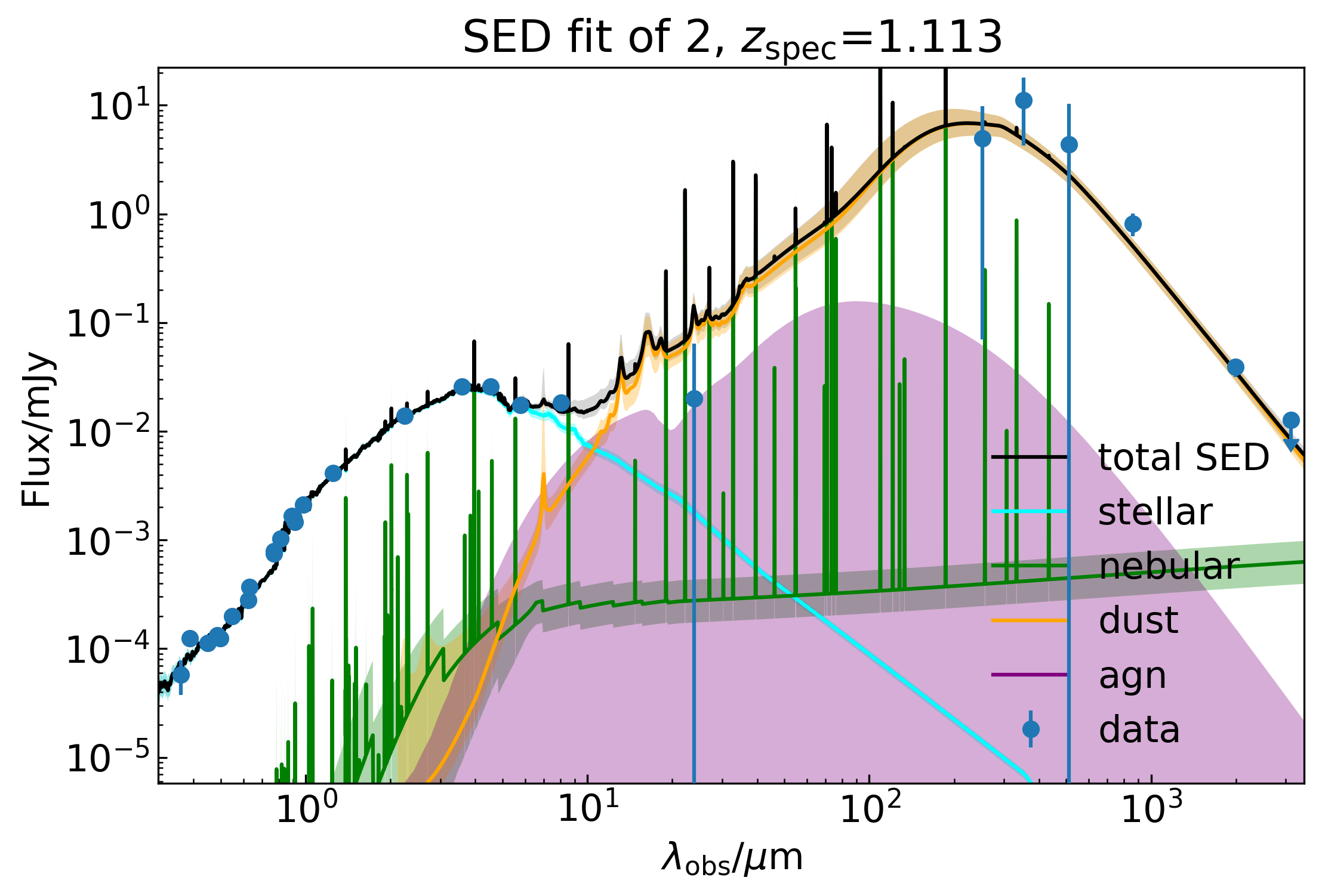}{0.5\textwidth}{(2)}}
    \gridline{\fig{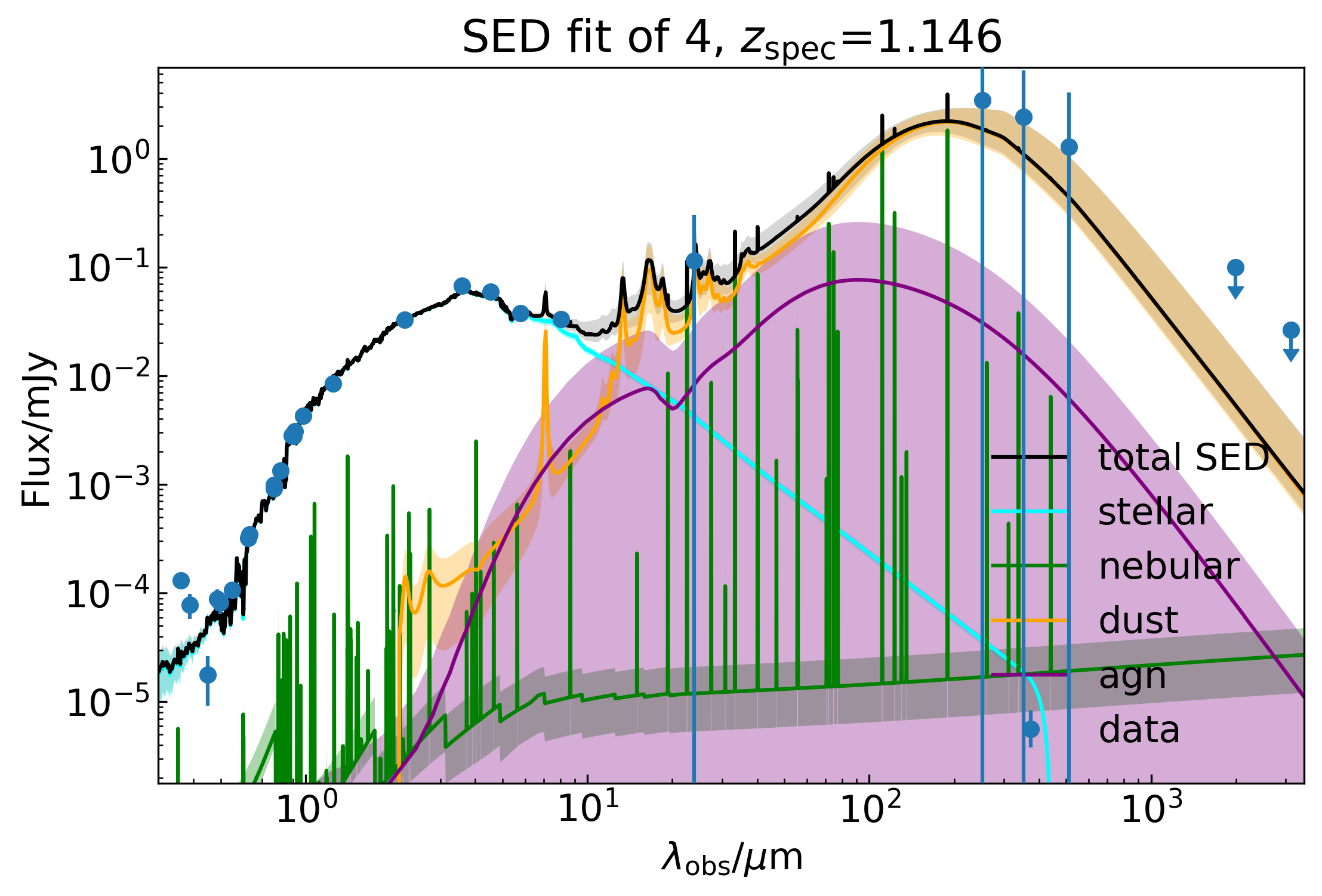}{0.5\textwidth}{(4)}}
    \caption{ Median (solid lines) and 16th-84th percentile range (shaded regions) SEDs of galaxies 1, 2, and 4, with individual components shown. The arrows indicate the ALMA $3\sigma$ flux upper limits used in the SED fitting. MIPS and Herschel fluxes from Bayesian deblending (Section \ref{subsubsec:irdeblending}) are directly fitted, regardless of S/N. The number of stacked ALMA sources with peak S/N$>2$ is less than 10 in the $u$ and NB359 bands, so the flux excesses of galaxy 4 in these bands might result from deblending issues due to poorly aligned astrometry.}
    \label{fig:figa1}
\end{figure*}

\begin{figure*}
\figurenum{A2}
    \gridline{\fig{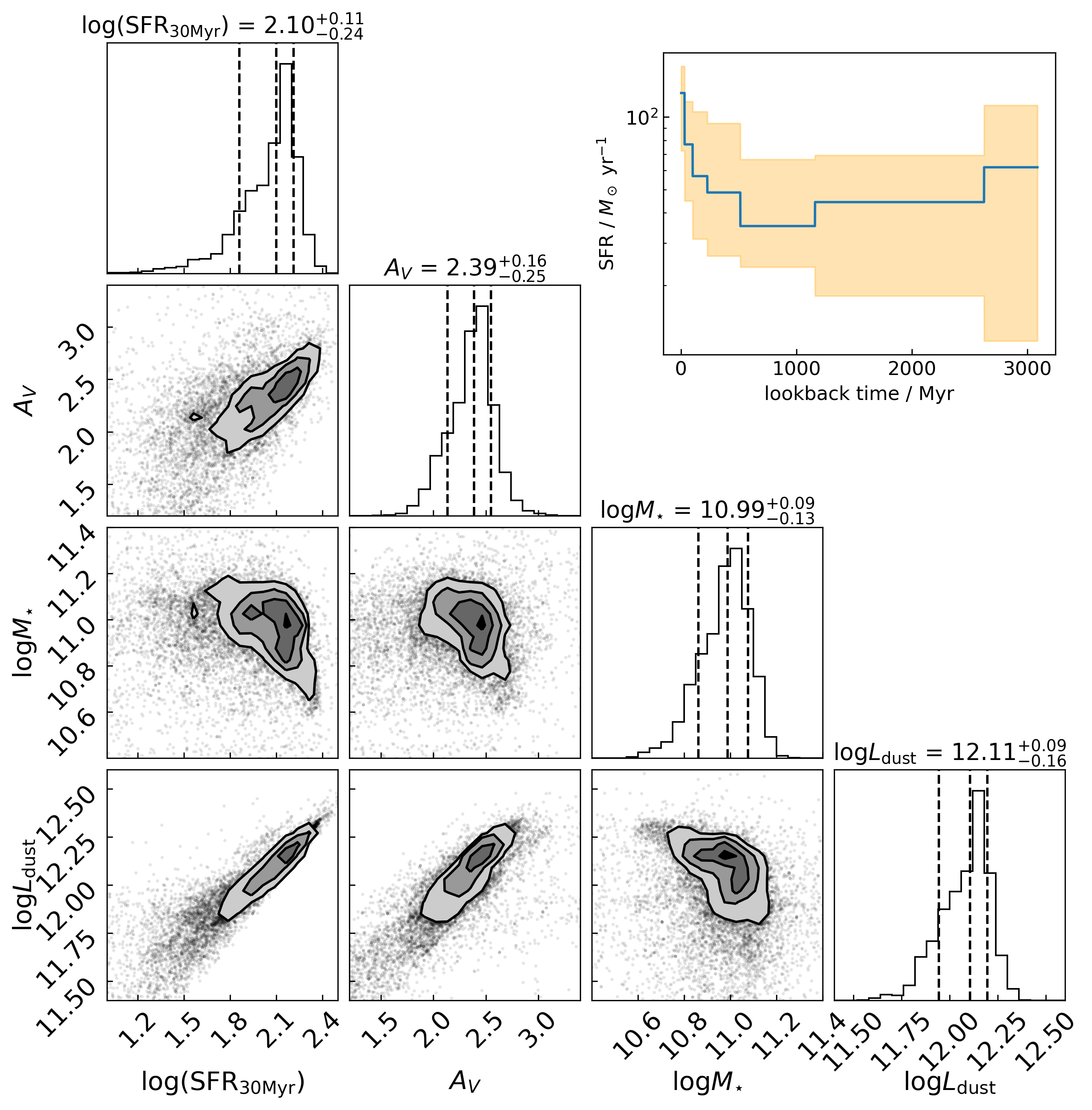}{0.5\textwidth}{(1)}\fig{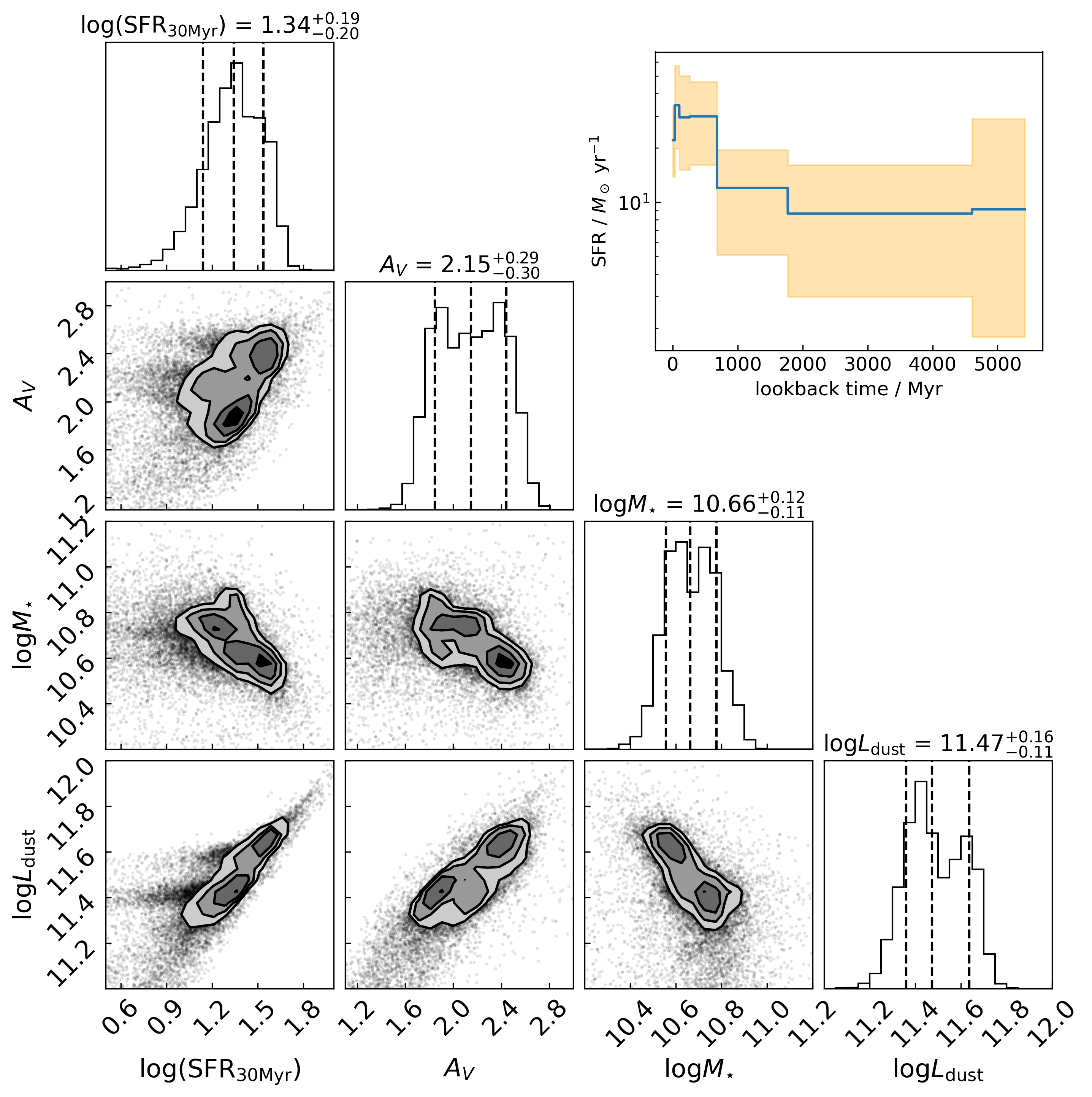}{0.5\textwidth}{(2)}}
    \gridline{\fig{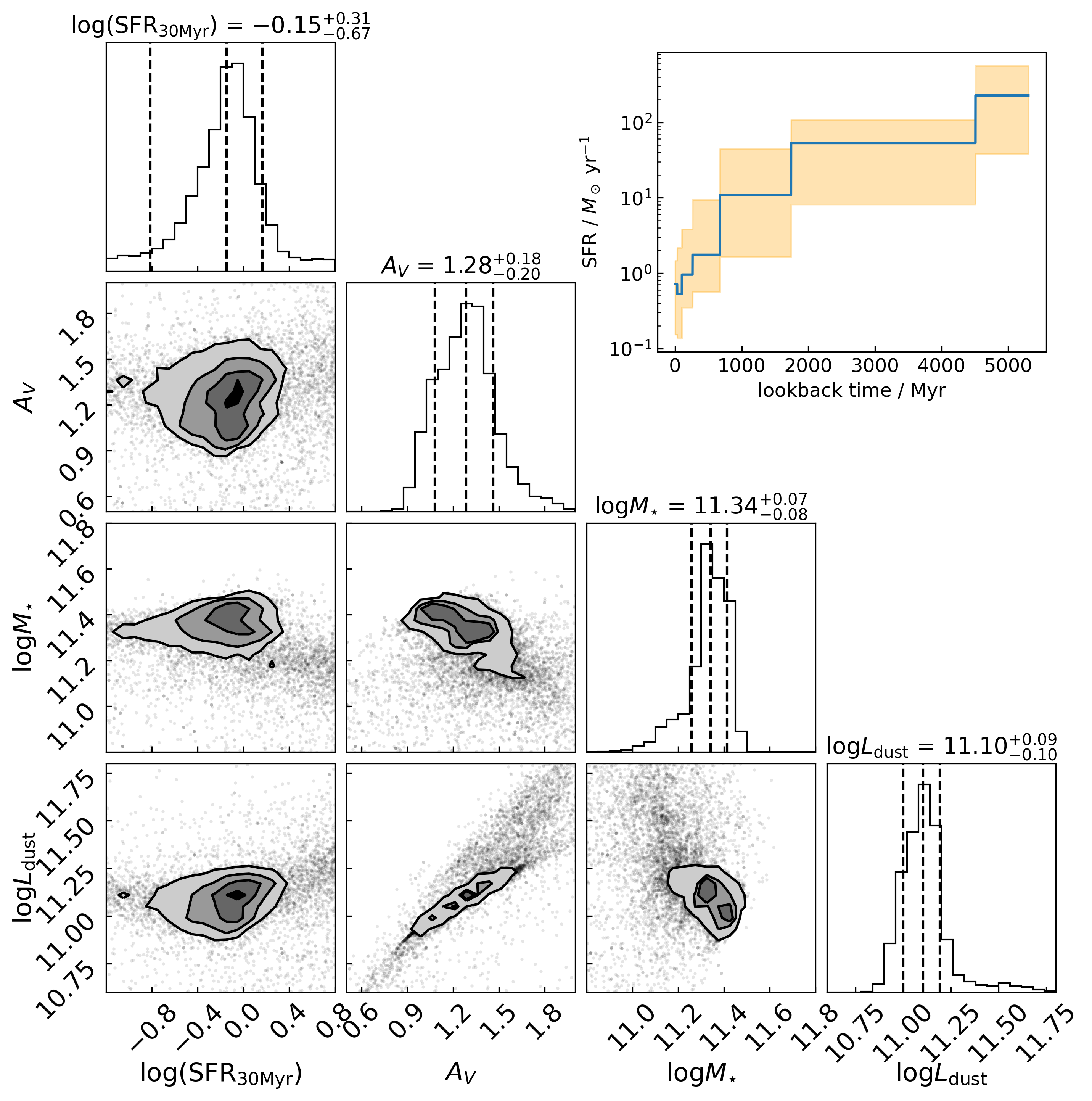}{0.5\textwidth}{(4)}}
    \caption{ Joint posterior distribution of derived physical parameters of galaxies 1, 2, and 4. Inserted panels show SFH from SED fitting with a blue solid line for the median and orange region for the 68\% percentile range.}
    \label{fig:figa2}
\end{figure*}

\bibliographystyle{aasjournal}
\bibliography{main}
\appendix
Figure \ref{fig:figa1} presents the SED fits of sources with CO-based redshift. Figure \ref{fig:figa2} presents the SFH and the joint posterior distribution of SED-derived physical parameters.
\section{SED fits}\label{appendixA}

\section{Posterior distribution of derived parameters}\label{appendixB}

\
\end{document}